\RequirePackage{fix-cm}
\documentclass[onecolumn]{svjour3}
\smartqed  %

\usepackage[square,sort,comma,numbers]{natbib}
\usepackage{booktabs} %
\usepackage[table]{xcolor}
\usepackage[flushleft]{threeparttable}
\usepackage{xcolor}
\usepackage[utf8]{inputenc}
\usepackage{float}
\usepackage{ifthen}
\usepackage{multirow}
\usepackage{caption}
\usepackage{listings}
\usepackage{hyperref}
\hypersetup{colorlinks=true}
\usepackage{url,moreverb,xspace}
\usepackage{tcolorbox}
\usepackage{enumitem}
\usepackage{array,graphicx}
\usepackage{soul}
\usepackage{balance}
\usepackage{pifont}
\usepackage{nicefrac}
\usepackage{mathtools}
\usepackage{amsmath,amssymb,amsfonts}
\usepackage{mathrsfs}
\usepackage{rotating}
\usepackage{ulem}  %

\usepackage[lined,boxruled,norelsize,linesnumbered]{algorithm2e}
\def\HiLi{\leavevmode\rlap{\hbox to \hsize{\color{gray!35}\leaders\hrule height .8\baselineskip depth .5ex\hfill}}}

\usepackage{tikz}
\usepackage{pgfplots}
\usepgfplotslibrary{statistics}
\usetikzlibrary{fadings}

\usepackage{xcolor,pifont}
\newcommand*\colourcheck[1]{%
  \expandafter\newcommand\csname #1check\endcsname{\textcolor{#1}{\ding{52}}\xspace}%
}
\newcommand*\colourcross[1]{%
  \expandafter\newcommand\csname #1cross\endcsname{\textcolor{#1}{\ding{56}}\xspace}%
}

\newboolean{showcomments}
\setboolean{showcomments}{true}
\ifthenelse{\boolean{showcomments}}
{
	\definecolor{myyellow}{RGB}{255, 228, 26}
	\definecolor{myblue}{RGB}{0, 0, 0}
	\definecolor{myblue2}{RGB}{50, 50, 220}
	\newcommand{\nb}[2]{
		{\sf
			\fcolorbox{myyellow}{yellow}{\scriptsize\textbf{#1}}%
			$\blacktriangleright$%
			{\color{myblue2}\fontsize{7pt}{8pt}\selectfont\textbf{#2}}%
		}%
	}
}
{
	\newcommand{\nb}[2]{}
}

\newcommand{\COMMENT}[1]{}

\newcommand{\gps}{GPSim\xspace}
\newcommand{\gpss}{GPSims\xspace} %
\newcommand{\hdt}{DT\xspace} %
\newcommand{\hdts}{DTs\xspace} %
\newcommand{\dk}{Donkey Car\xspace} %
\newcommand{\davetwo}{\mbox{Dave-2}\xspace} %
\newcommand{\chauffeur}{\mbox{Chauffeur}\xspace} %
\newcommand{\epoch}{\mbox{Epoch}\xspace} %
\newcommand{\msim}{$M_S$\xspace} %
\newcommand{\mreal}{$M_R$\xspace} %
\newcommand{\dsone}{DS\textsubscript{1}\xspace} %
\newcommand{\dstwo}{DS\textsubscript{2}\xspace} %
\newcommand{\dss}{DSS\xspace} %

\newcommand{\head}[1]{\noindent\textbf{#1.}}

\makeatletter
\let\orgdescriptionlabel\descriptionlabel
\renewcommand*{\descriptionlabel}[1]{%
  \let\orglabel\label
  \let\label\@gobble
  \phantomsection
  \edef\@currentlabel{#1}%
  \let\label\orglabel
  \orgdescriptionlabel{#1}%
}
\makeatother

\journalname{EMSE}

\begin{document}

\title{
Two is Better Than One: Digital Siblings to Improve Autonomous Driving Testing
}

\author{Matteo Biagiola \and
	Andrea Stocco \and
	Vincenzo Riccio \and
	Paolo Tonella
}

\authorrunning{Biagiola M., Stocco A., Riccio V. and Tonella P.}%
\titlerunning{Two is Better Than One: Digital Siblings to Improve Autonomous Driving Testing}%

\institute{M. Biagiola and P. Tonella \at
	Universit\`a della Svizzera italiana (USI),
	Via Buffi, 13 -- Lugano, Switzerland \\
	tel +41 58 666 40 00, 
	fax +41 58 666 46 47 \\
	\email{\{matteo.biagiola,paolo.tonella\}@usi.ch} \\
	A. Stocco \at 
	Technical University of Munich -- Boltzmannstra{\ss}e 3 Garching near Munich, Germany and fortiss GmbH -- Guerickestra{\ss}e 25 Munich, Germany \\
	\email{andrea.stocco@tum.de|stocco@fortiss.org} \\
	V. Riccio \at
	Universit\`a degli Studi di Udine, 
	Via Gemona 92 -- Udine, Italy \\
	tel +39 0432 556680 \\
	\email{vincenzo.riccio@uniud.it}
}

\maketitle

\begin{abstract}
Simulation-based testing represents an important step to ensure the reliability of autonomous driving software. In practice, when companies rely on third-party general-purpose simulators, either for in-house or outsourced testing, the generalizability of testing results to real autonomous vehicles is at stake. 
In this paper, we enhance simulation-based testing by introducing the notion of \textit{digital siblings}---a multi-simulator approach that tests a given autonomous vehicle on multiple general-purpose simulators built with different technologies, that operate collectively as an ensemble in the testing process.

We exemplify our approach on a case study focused on testing the lane-keeping component of an autonomous vehicle. We use two open-source simulators as digital siblings, and we empirically compare such a multi-simulator approach against a digital twin of a physical scaled autonomous vehicle on a large set of test cases. 
Our approach requires generating and running test cases for each individual simulator, in the form of sequences of road points. Then, test cases are migrated between simulators, using feature maps to characterize the exercised driving conditions. Finally, the joint predicted failure probability is computed, and a failure is reported only in cases of agreement among the siblings.

Our empirical evaluation shows that the ensemble failure predictor by the digital siblings is superior to each individual simulator at predicting the failures of the digital twin. We discuss the findings of our case study and detail how our approach can help researchers interested in automated testing of autonomous driving software.

\end{abstract}


\section{Introduction}\label{sec:introduction}

The development of autonomous vehicles (AVs) has received great attention in the last decade. As of 2020, more than \$150 billions have been invested in AVs, a sum that is expected to double in the near future~\cite{ad-market}. 
AVs typically integrate multiple advanced driver-assistance systems (e.g., for adaptive cruise control, parking assistance, and lane-keeping) into a unified control unit, using a perception-plan-execution strategy~\cite{yurtsever2020survey}. 
Advanced driver-assistance systems based on Deep Neural Networks (DNNs) are trained on labeled input-output samples of real-world driving data provided by the vehicle sensory to learn human-like driving actions~\cite{grigorescu2020survey}.

Before deployment on public roads, AVs are thoroughly tested in the field, on private test tracks~\cite{waymo,borg,comprehensive-sfc-test,stocco-mind}. While essential for fully assessing the dependability of AVs on the road, field testing has known limitations in terms of cost, safety and adequacy~\cite{stocco-mind}. 
To overcome these limitations, driving simulators are used to generate several real-life edge case scenarios that are unlikely to be experienced during field testing, or that are dangerous to reproduce for human operators~\cite{borg,challenges-av-testing}.
Simulation-based testing represents a consolidated testing practice, being more affordable than field testing, yet capable of exposing many bugs before deployment~\cite{waymo,borg,comprehensive-sfc-test,stocco-mind}. 

In this paper, we distinguish two main categories of driving simulators, namely digital twins (\hdts) and general-purpose simulators (\gpss).
\hdts provide a software replica of \textit{specific} real vehicles, that are digitally recreated in terms of appearance, aerodynamics, and physical interactions with the environment~\cite{borg}. In the context of mixed-reality testing approaches~\cite{survey-lei-ma,nhtsa}, such as Hardware-in-the-Loop and Vehicle-in-the-Loop, the digital twin is connected to physical AV components to further increase the degree of fidelity. In this paper, we consider simulation-based testing where the digital twin is a software replica of a specific real vehicle. Developing a \hdt is expensive~\cite{siemens,cagatay} and can take up to five years~\cite{wayve-infinity}. Hence, it remains an exclusive prerogative of big companies such as Uber (Waabi World~\cite{waabi-world}), Waymo (Simulation City~\cite{simulation-city}) or Wayve (Infinity Simulator~\cite{wayve-infinity}).
\gps are generally designed without the need to faithfully reproduce a specific vehicle or testing scenario, as they rather offer generic APIs to run one or more AVs on virtual road tracks. \gps such as Siemens PreScan~\cite{prescan} or ESI Pro-SiVIC~\cite{pro-sivic} offer a more affordable alternative to the expensive \hdt development, and are widely used for outsourcing testing tasks to third-party companies~\cite{outsourcing-av-development}, for which access to, or customizations of the original \hdt are not feasible for each individual vehicle~\cite{hu2023sim2real}. 

Despite affordability, \gps can be affected by a \textit{fidelity} and \textit{reality gap}, when the simulated experience does not successfully transfer from the \gps to the reference \hdt and eventually to the real AV~\cite{hu2023sim2real}. 
These discrepancies can lead to a distrust in simulation-based testing, as reported by recent surveys~\cite{icst-survey-robotics,fse-survey-robotics,hu2023sim2real,survey-lei-ma}. 
While comparative works of \gps exist in the literature~\cite{DBLP:journals/corr/abs-2101-05337,s19030648}, cross-simulator testing for AVs is a relatively unexplored avenue for research. Only a recent study~\cite{borg} investigates the use of multiple \gps for  testing a pedestrian vision detection system. The study compares a large set of test scenarios on both PreScan~\cite{prescan} and Pro-SiVIC~\cite{pro-sivic} and reports inconsistent results in terms of safety violations and behaviors across these simulators. Consequently, using a single-simulator approach for AV testing might be unreliable, as the testing results are highly dependent on the chosen \gps.

In this paper, we target the fidelity gap between \gps and \hdt by proposing a multi-simulator approach for AV testing called \textit{digital siblings} (\dss).
Our approach involves automated test generation and a novel cross-simulator feature map analysis that combines the outcome of several simulator-specific test generators into a unified view. We use \dss as a surrogate model of the behavior of a \hdt. Our intuition is that agreement among multiple \gps will increase the confidence in observing the same behavior in \hdt. On the other hand, in the presence of disagreements, \dss can mitigate or even eliminate the risk of choosing the worst \gps, which would give poor simulation testing results.

In detail, our case study consists in the automatic generation of test cases, i.e., sequences of road points determining the roads where the AV drives, to test the lane-keeping component of an AV. We then use feature maps to characterize both the structure of such test cases, and the behaviors of the AV in each of them, to group failures by similarity, and to avoid reporting the same failures repeatedly. 
To account for the specificities of each \gps, we execute test generation separately for each sibling. Then, we migrate the tests generated for one sibling to the other sibling. Finally, we merge failing and non failing executions based on similarity of features and estimate the overall joint failure probability. 

In our case study we use \dss to test three state-of-the-art DNN lane-keeping models, i.e., Nvidia \davetwo~\cite{nvidia-dave2}, Chauffeur~\cite{chauffeur}, and Epoch~\cite{epoch} (the last two were developed by the respective teams in the Udacity challenge competition~\cite{udacity-challenge}). We consider as siblings two open-source simulators, namely Udacity~\cite{udacity-sim} and BeamNG~\cite{beamng}, widely used in previous studies to test lane-keeping software~\cite{asfault,2021-Jahangirova-ICST,deepjanus,2020-Stocco-ICSE,deephyperion}. 
As \hdt, we adopt an open-source framework~\cite{donkey} used in previous research~\cite{stocco-mind,survey-lei-ma,2021-01-0248,viitala,9412011} featuring a virtual replica of a 1:16 scale electric AV. 
We evaluate \dss with both \textit{offline} and \textit{online} testing~\cite{briand-offline-emse}, i.e., the lane-keeping models are tested both w.r.t. the accuracy of its predictions on labeled individual inputs, and at the system-level for their capability to control the vehicle on several hundreds automatically-generated roads.  

Our empirical evaluation shows that, at the model-level, the distribution of prediction errors of \dss is statistically indistinguishable from that of the \hdt.
Overall, at the system-level, the failure probability of \dss highly correlates with the true failure probability of the \hdt.
More notably, the quality of driving measured in \dss can predict the true failure probability of the \hdt, which suggests that we can use the digital siblings to possibly anticipate the failures of the lane-keeping component of the real-world AV more reliably than with a single \gps. 
A practical implication of our findings for software engineers is the usage of digital siblings when testing DNN-based lane-keeping software, to increase the level of fidelity of the observed behaviors and failures. The same recommendation holds for AV testing researchers.

\clearpage

Our paper makes the following contributions:
\begin{itemize}
\item \textbf{Digital Siblings.} A novel approach for testing DNN-based lane-keeping software that generates road scenarios in multiple general-purpose simulators, and combines their testing outcomes to approximate a digital twin. This is the first solution that leverages a multi-simulator approach to overcome the simulation fidelity gap. 
\item \textbf{Evaluation.} An empirical study showing that the digital siblings are effective at predicting the failures of the AV under test in the digital twin for a physical scaled vehicle in the lane-keeping task. 
\end{itemize}

\section{Motivation and Background} \label{sec:motivation-and-background}

In this section, we provide additional motivation for our approach, and we briefly describe the main concepts to understand the rest of the paper. In particular, we discuss the lane-keeping functionality of an AV, and we introduce evolutionary search as a tool to generate challenging test scenarios for AVs.

\subsection{Motivation}

In practice, test engineers use simulation platforms for testing early releases of their autonomous driving software, prior to real-world physical testing. 
The gap between simulated and real-world test outcomes hinders trustworthiness in the testing process. Thus, efforts must be made to provide evidence that simulation-based testing campaigns can expose real-world AV failures.

In an ideal scenario, the chosen simulation platform is able to accurately replicate the physics of the AV under test. Such high-fidelity digital twins are used by automotive companies as a proxy for their physical AVs. Under this assumption, the high-fidelity digital twin allows to safely carry out a testing campaign while saving costs and, at the same time, improving the robustness of the software.

However, high-fidelity digital twins are costly to develop and maintain, and not all manufacturers can afford them (those who can are not keen to disclose their high-fidelity digital twins, as these are valuable assets that give them a competitive advantage). Moreover, AV manufacturers outsource most of the testing processes to small/medium companies and such high-fidelity digital twins are not available to them. These companies adopt \gpss as a low-cost alternative for simulation-based testing of AVs.

The goal of our approach is to increase the \textit{reliability} of simulation-based testing of AVs, specifically targeting environments that adopt general purpose simulators that are not designed to represent a specific AV, but rather focus on high-level scenario-based testing. 
To mitigate this design limitation, we propose a testing methodology employing an ensemble of \gpss. This approach involves aggregating the outcomes of multiple \gpss to mitigate the risks associated with simulator flakiness or representativeness. 
We combine multiple relatively low-cost simulators to obtain reliable test results as if we used a very costly dataset from the real operation or a high-cost simulator such as a high-fidelity digital twin.
Our approach is particularly beneficial when these \gpss exhibit complementary behaviors, allowing them to compensate for each other's weaknesses while combining their strengths. 
Our research hypothesis is that the combination of complementary \gpss provides a more reliable estimation of testing outcomes than the usage of a single \gps.
In this paper, we present the initial findings supporting this hypothesis, exploring and evaluating one practical implementation of our approach using widely accessible open-source simulation platforms.

We instantiate our approach for testing the lane-keeping component of an AV, implemented with a DNN. The test cases are sequences of road points, which determine the two-lane roads where the AV is supposed to drive autonomously. To assess the benefits of our multi-simulator approach (i.e., \dss), we use the digital twin (\hdt) of a physical 1:16 scale electric AV~\cite{donkey}, as a surrogate for the real-world AV behaviors. Indeed, we assume having access only to multiple \gpss as, in practice, a \hdt is often unavailable. In our evaluation, we validate our hypothesis by comparing the extent to which both \dss and each individual sibling can predict the failures of the DNN lane-keeping component in \hdt, thus quantifying the reliability of testing.

\subsection{Background}

\subsubsection{Lane-keeping}

This paper focuses on testing AVs that perform the lane-keeping functionality from driving samples labeled by humans. 
Lane-keeping, also called lane-centering or lane-following, is an automated driving assistance feature of an AV to keep the vehicle at the center of the lane. This system can be implemented at different levels, from a warning to the driver when the vehicle crosses one of the lanes up to the driverless version, which steers the vehicle automatically when it detects a departure from the center of the lane.

In this paper we consider the driverless version since it is a crucial component for the safe deployment of AVs on public roads. Indeed, according to a report by NHTSA~\cite{precrashreport}, off-road crashes due to failures of the lane-keeping component are first in cost (\$15 billion) and second in frequency.
From a technical standpoint, the lane-keeping task is implemented by \textit{behavior cloning} DNNs, which learn end-to-end from supervised expert demonstrations. The training dataset consists of driving images captured with a camera sensor mounted on board of the vehicle, appropriately labeled with the driving commands of a human driver.

We consider lane-keeping DNN models, such as NVIDIA's \davetwo~\cite{nvidia-dave2}, that predict the steering angle at which the car should steer to keep the vehicle in lane, given a single driving image. These models are generally trained with stochastic gradient descent~\cite{saadonline} on stationary datasets, with the goal of minimizing the error between the predicted and the ground-truth steering angles.

Such labels are typically an array of commands, i.e., steering, throttle and brake, although in the simplest case only the steering is provided, while the throttle is determined as a function of the steering and the velocity of the vehicle. 
Given the dataset, a DNN model, such as the \davetwo model from Nvidia~\cite{nvidia-dave2}, is trained to predict the label given an image by minimizing the Mean Squared Error (MSE) between the current prediction and the ground-truth label.

\subsubsection{Evolutionary Search} 

Evolutionary or metaheuristic search is a class of techniques that apply randomness and heuristics to find near-optimal solutions to optimization problems~\cite{essentials}. Such techniques are very general, since they only require evaluating how good a candidate solution is. The \textit{goodness} of a solution is called fitness and the objective of the search algorithm is to optimize it (either maximize it or minimize it). The algorithm manipulates a solution to exploit the known parts of the search space, and creates new solutions to explore the parts that are unknown.

Search algorithms have been applied to testing problems and have been particularly effective tools for test generation~\cite{evosuite,dynamosa,pynguin}. In this paper, we use the MapElites search algorithm~\cite{mapelites}, implemented in the DeepHyperion tool~\cite{deephyperion}, to generate test cases for the DNN model under test. The algorithm explores the solution feature space at large, in order to provide a comprehensive characterization of the behaviors of the driving model.

\section{Multi-simulator AV Testing with Digital Siblings} \label{sec:approach}

\begin{figure*}[t]
\centering

\includegraphics[trim=1cm 10cm 22cm 1.5cm, clip=true, scale=0.13]
{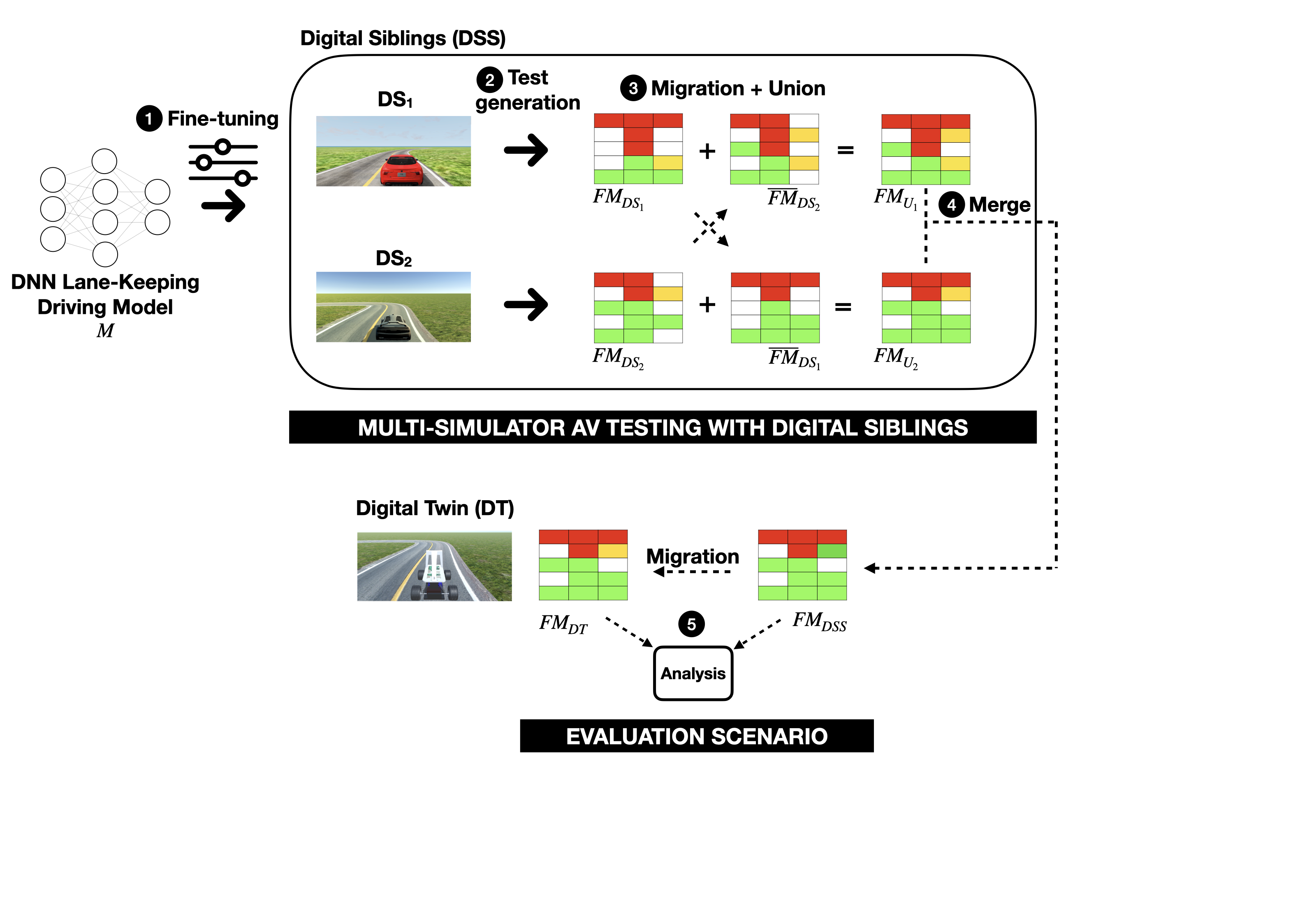}

\caption{Overview of our multi-simulator approach and its usage.} 
\label{fig:approach:overview} 
\end{figure*}

The goal of our approach is to use digital siblings to test the DNN-based lane-keeping component of an AV, by generating a large set of road scenarios. Our approach takes as input a DNN lane-keeping model $M$, and uses an existing road generator to test its behavior, by generating roads for multiple driving simulators. The key intuition is that multiple \gpss can better approximate the driving behavior of the AV executed in \hdt, which we use as a proxy for the behavior of the real-world AV, as opposed to a single-simulator approach.

Our approach supports an arbitrary number of digital siblings. 
For simplicity of exposition, engineering effort, and evaluation, we describe and expe\-riment it using two simulators. However, we present the most important steps of our approach, i.e., migration (step~\ding{184}) and merge (step~\ding{185}), in a generic manner that accommodates any number of siblings.

\autoref{fig:approach:overview} (top) shows an overview of our approach in which two digital siblings, namely \dsone and \dstwo, are used to test the behavior of a driving model under test $M$, i.e., an end-to-end DNN for lane-keeping. 
In the first phase, $M$ is either trained or fine-tuned (step~\ding{182}) to run on both \dsone and \dstwo, as well as on the target platform (i.e., \hdt).
A test generation phase (step~\ding{183}) is executed for each digital sibling, generating road scenarios for each simulator and producing two \textit{feature maps} $FM_{DS_1}$ and $FM_{DS_2}$.
Feature maps group together test cases with similar feature combination values, to reduce redundancy and summarize the AV behaviors in unique feature combination~\cite{deephyperion,zohdinasabefficient}.
The value in a feature map cell, displayed in a colored heat scale, represents the average test case outcome, i.e., the behavioral information about the execution of $M$ in each test scenario (e.g., the failure probability).  
For each simulator, the test generation algorithm produces test scenarios that are executed to assess the behavior of the driving model $M$ under many different circumstances. Hence, the output of test generation is simulator and model dependent and the feature maps of \dsone ($FM_{DS_1}$) and \dstwo ($FM_{DS_2}$) can be different. 

The next step of our approach (step~\ding{184}) requires to \textit{migrate} the test cases across simulators. In detail, the test cases in $FM_{DS_1}$ are executed on \dstwo, resulting in the feature map $\overline{FM}_{DS_1}$. Similarly, the test cases in $FM_{DS_2}$ are executed on \dsone, resulting in the feature map $\overline{FM}_{DS_2}$. Then, for both \dsone and \dstwo, we compute the \textit{union} of the two feature maps, obtaining $FM_{U_1}$ for \dsone and $FM_{U_2}$ for \dstwo. Both maps contain the same set of test cases, although executed on two different simulators. The final output of the digital siblings (step~\ding{185}) is obtained by \textit{merging} $FM_{U_1}$ and $FM_{U_2}$ into the final feature map $FM_{DSS}$. 

Step~\ding{186} assesses the correlation of the $FM_{DSS}$ map with the $FM_{\hdt}$ map, to evaluate the predictive capability of the digital siblings. 
\autoref{fig:approach:overview} (bottom) shows an overview of the evaluation of our approach (detailed later, in \autoref{sec:study}).  
All the test cases in the final feature map $FM_{DSS}$ are executed (i.e., migrated) on the \hdt, to obtain the ground truth feature map $FM_{\hdt}$. 

\subsection{Test Scenarios}

\begin{figure}[t]
\centering

\includegraphics[trim=0cm 26cm 44cm 0cm, clip=true, scale=0.25]
{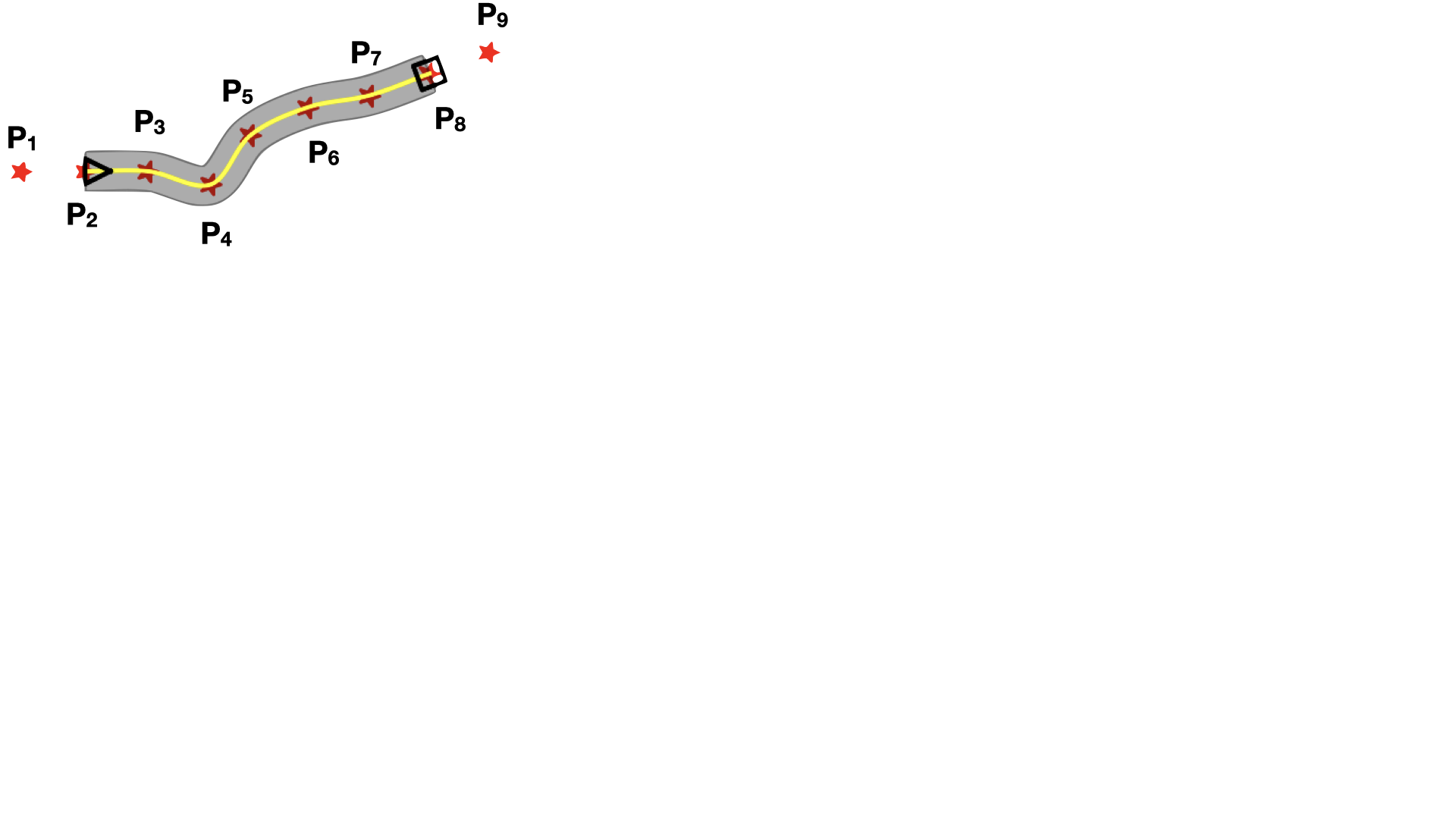}

\caption{Example of test scenario for a lane-keeping autonomous driving system.} 
\label{fig:approach:road-segment} 
\end{figure}

\subsubsection{Representation}

We adopted an abstract representation of the road in each driving simulator so that only a sequence of road control points is needed when creating a new road in the driving scene. 
We follow the representation given by Riccio and Tonella~\cite{deepjanus} who defined a two-lane road using a series of \textit{control points} (displayed as red stars in \autoref{fig:approach:road-segment}). The control points are interpolated using \textit{Catmull-Rom} splines~\cite{10.1145/378456.378511}, giving the road its final shape (yellow solid line). 

\autoref{fig:approach:road-segment} shows the visualization of a test scenario generated at step~\ding{183}. 
Specifically, the road is defined using nine control points whereas the Catmull-Rom spline only goes through seven of them. This is because a spline segment (e.g., $P_2-P_3$) is always defined by four control points (e.g., $P_1$, $P_2$, $P_3$, $P_4$). Since two of them are on either side of the endpoints of the spline segment (e.g., $P_1$ and $P_4$), the spline cannot traverse the extreme endpoints (e.g., $P_1$ and $P_9$). Hence, $P_2$ defines the start point of the road (depicted as a black triangle) whereas $P_8$ defines the end point (depicted as a black square). 

\subsubsection{Implementation}

The default initial state of each test case involves positioning the vehicle in the first drivable control point (i.e., $P_2$ in \autoref{fig:approach:road-segment}), at the center of the right lane following the road orientation.

We uniformed the 3D rendering of each simulator such that the driving scenarios have the same look and feel: a two-lane asphalt road, where the road is delimited by two solid white lines on each side and the two driving lanes are separated by a single solid yellow line. The road is placed on top of a green plane representing grass. 
Harmonization of the driving scenarios across simulators ensures that geometrical features are preserved for the collected driving images and that any color transformation applied to them during training preprocessing remains applicable~\cite{nvidia-dave2}.

\subsubsection{Validity and Oracle}\label{sec:oracle}

After interpolation, a road is deemed \textit{valid} if it respects the following constraints: (1)~the start and end points are different; (2)~the road is contained within a squared bounding box of a predefined size (specifically 250 $\times$ 250); and, (3)~there are no intersections. 
A test case is deemed \textit{successful} when the vehicle drives within the right lane until the last road control point (e.g., $P_8$ in \autoref{fig:approach:road-segment}). On the contrary, a test case \textit{failure} occurs when the vehicle drives \textit{out of bound} (OOB).

\subsection{Creating/Fine-Tuning the Driving Model}

\subsubsection{Data Collection} 
For the creation or fine-tuning of a self-driving model (step~\ding{182}), a labeled dataset of driving scenes is needed. 
We automate labeled data collection by resorting to \textit{autopilots} that have \textit{global knowledge} of the driving scenario such as the detailed road geometry and precise vehicle position. 
In particular, in each simulator, at each step of the simulation, the steering angle of the autopilot is computed by a Proportional-Integral-Differential (PID) controller~\cite{farag2020complex} according to the formula:

\begin{equation} \label{eq:approach:pid}
	\textit{steering} = K_P \cdot \textsc{LP} + K_D \cdot \textit{diff}_{\textsc{LP}} + K_I \cdot \textit{total}_{\textsc{LP}}
\end{equation}

\noindent
where LP stands for \textit{lateral position}~\cite{2020-Stocco-GAUSS} (in particular, the lateral position is zero when the vehicle drives at the center of the lane). \autoref{eq:approach:pid} states that the proportional constant $K_P$ acts on the raw error while the derivative constant $K_D$ controls the difference between two consecutive errors and the integral constant $K_I$ considers the total sum of the errors during the whole simulation until the current timestep. Finally, the steering value is clipped in the interval $[-1, +1]$, where $-1$ means steering all the way to the left and $+1$ to the right ($0$ means the vehicle goes straight as no steering is applied). The steering values are normalized in order to account for the different simulators that we use in our approach.

The autopilot produces a steering angle label for each image which is used to train the driving model. We aligned the frame rates of the different simulators at 20 \textit{fps} such that, in each simulator, the autopilot collects a comparable number of labeled images. 
The speed of the vehicle, both for the autopilot and $M$, is controlled by the throttle via a linear interpolation between the minimum speed and maximum speed so that the car decreases the speed when the steering angle increases (e.g., in a curve). The following formula computes the throttle based on the speed of the vehicle and the steering:

\begin{equation} \label{eq:approach:throttle}
	\textit{throttle} = 1 - \textit{steering}^2 - \Big(\frac{\textit{speed}}{\textit{K}}\Big)^2
\end{equation}

\noindent
where $K$ is set to a predefined low  value $L$ when the measured \textit{speed} is greater than a given maximum speed threshold, to enforce strong deceleration; viceversa, $K$ is set to a high value $H$ when the measured \textit{speed} is lower than or equal to the maximum speed threshold, to reduce the deceleration component. From \autoref{eq:approach:throttle}, we can see that the throttle is close to 1 (the highest possible value) when the vehicle does not steer ($\textit{steering} = 0$) and the \textit{speed} is substantially lower than the maximum allowed speed (in this case, $K=H$); when one of the two conditions is false the throttle decreases, because of either deceleration component. Similarly to the steering angle values, we clip the throttle value in the interval $[0, 1]$.

\subsubsection{Model Fine-Tuning via Hybrid Training} 
The next step involves training the model $M$ using all simulators and the data collected in step~\ding{182}. Alternatively, if an existing trained model $M$ is available for the target \hdt, our approach requires \textit{fine-tuning} it for all digital siblings. 
In both scenarios, we use \textit{hybrid} training based on gradient descent~\cite{10.5555/2981562.2981583}. 

Hybrid training requires combining the  datasets collected for different simulators/platforms into a unified dataset, making sure that each dataset is equally represented (i.e., the unified dataset contains the same number of samples from each simulator/platform specific dataset). Then, the unified dataset is split into training and validation sets (e.g.,  using the standard 80/20 ratio).
The training pipeline is designed in such a way that each image, of dimensions 320$\times$160, is processed according to the simulator/platform it was taken from. For example,  images may be cropped differently. Depending on the vehicle size, the front part of the car may, or may not be visible in the frame captured by the camera. Another example of simulator-specific adaptation is the cropping of the above-horizon portion of the image, unnecessary for the lane-keeping task.
After cropping, each image is resized to the size required for training, i.e., 320$\times$160.

The training pipeline can be further configured to use plain synthetic virtual images from the driving simulators, or pseudo-real images resembling real-world driving images.
The first configuration represents the standard practice in AV testing. 
In the second configuration, the reality gap due to low photo-realism is reduced by an \textit{image-to-image} transformation that translates the driving images of each simulator into images similar to those captured by the real-world AV during on-road driving. This practice was proposed in the literature~\cite{stocco-mind} and in industry~\cite{wayve-sim2real} to increase the transferability of the driving model tested in simulation to the real world. 

More specifically, this second configuration requires training a CycleGAN model for each driving simulator~\cite{cyclegan}. CycleGAN entails two \textit{generators}, one that learns how to translate images from \textit{simulated} to \textit{real} world (sim2real) and the other that learns the opposite transformation (real2sim). During training of the model, we use the sim2real generator trained for the respective simulator to translate the corresponding training set images. 
During testing, the sim2real generator translates images at runtime, i.e., during the execution of the simulation. We refer to the translated images as \textit{pseudo}-real, since they are the output of a generative process designed to resemble real images.

\begin{figure}[t]
\centering

\includegraphics[trim=0cm 25cm 4cm 0cm, clip=true, scale=0.16]
{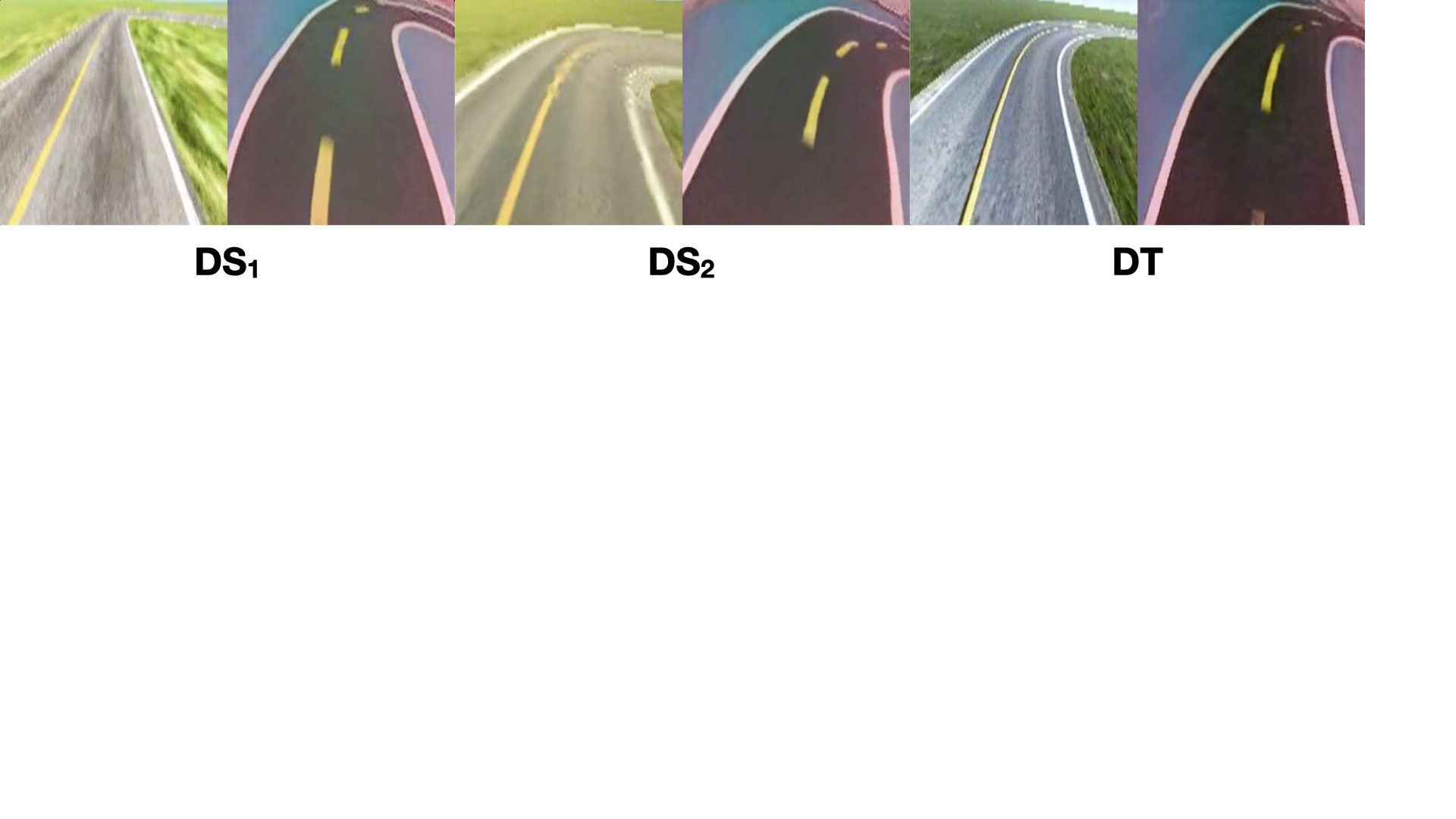}

\caption{Example of CycleGAN translation for the three simulators.} 
\label{fig:approach:image-translation} 
\end{figure}

\autoref{fig:approach:image-translation} shows an example of image translation with a CycleGAN trained for each simulator. The corresponding networks translate an image of a road curve taken in the simulated domain (left) to an image belonging to the real domain (right)---the test track of a small scale physical AV. 
During training and testing of the driving model in a given simulator, we use the generator of the CycleGAN trained for such simulator.

In our evaluation (\autoref{sec:study}), we consider both configurations of our approach, i.e., training using either simulator or pseudo-real images. We refer to the model trained on simulator images as \msim, and the model trained on pseudo-real images as \mreal.

\begin{figure}[b]
\centering

\includegraphics[trim=2cm 28cm 10cm 0cm, clip=true, scale=0.12]
{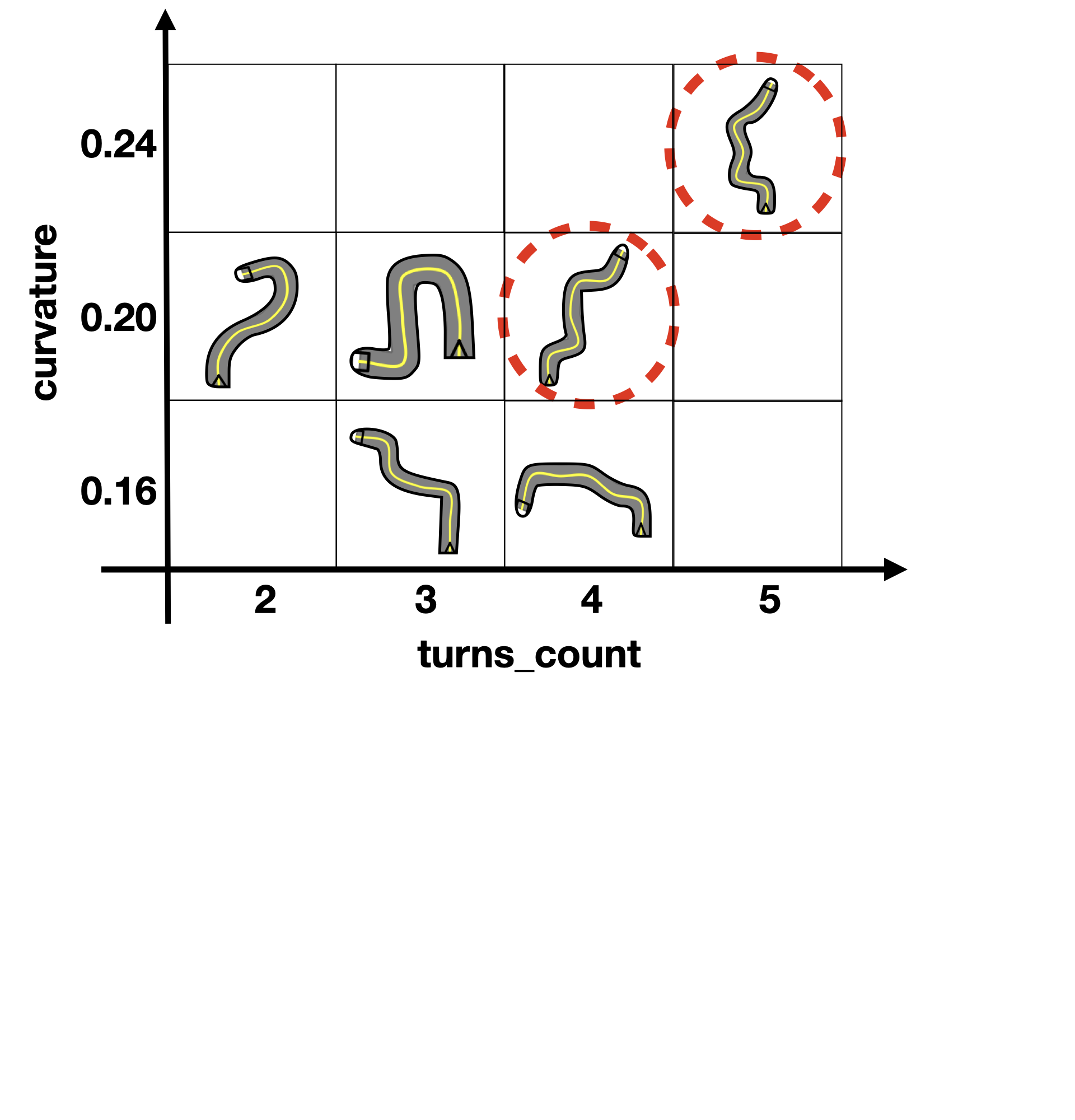}

\caption{Example of feature map by DeepHyperion. The two axes represent structural features of the generated roads (i.e., curvature and number of bends).} 
\label{fig:approach:feature-map} 
\end{figure}

\subsection{Test Generation}

While our approach is compatible with any test generation algorithm, in this paper we adopt the \textit{MapElites}~\cite{mapelites} algorithm implemented in DeepHyperion~\cite{deephyperion}, because
the output of DeepHyperion is projected to a feature map that characterizes each generated test scenario according to its features. In other words, test cases having equivalent features (e.g., 3 turns and maximum curvature of 0.2) are grouped into the same \textit{cell} of the feature map.

\autoref{fig:approach:feature-map} shows an example of feature map generated by DeepHyperion. The roads (i.e., the test cases) in the map are characterized by two structural features, i.e., the \textit{number of turns} in the road ($x$ axis) and the \textit{curvature} of the road ($y$ axis), the latter defined as the minimum radius of the circles going through each sequence of three consecutive road points~\cite{deephyperion}. Such features have been used in previous work and have been shown to be effective at characterizing the search space of road generators~\cite{deephyperion}. Characterizing a test case based on its structural features, i.e., only based on the properties of the road, allows us to identify unique failure scenarios, i.e., failure scenarios with distinctive road properties.

During test generation, the test cases are distributed in the map according to their features. The \textit{value} of each cell is influenced by the behavior of $M$ when driving on the roads pertaining to a cell. The minimum \textit{lateral distance} recorded by the simulator is used by DeepHyperion as a \textit{fitness} of the generated test case. The lateral distance is the opposite of the lateral position, i.e., it has the highest value when the vehicle drives at the center of the lane, and it decreases as the vehicle approaches the roadside. In particular, it is negative when the model misbehaves (i.e., the vehicle goes out of bound). In \autoref{fig:approach:feature-map} the two dashed-encircled cells point out two failure cells for $M$ (i.e., cells containing roads with negative fitness).

\autoref{algorithm:approach:deephyperion} shows the pseudocode of the DeepHyperion algorithm. It takes as input the driving model under test $M$, the simulator instance $S$ and two hyperparameters, i.e., the population size $P_s$ and the number of iterations $N$ the search is allowed to run, i.e., the budget of the algorithm. The algorithm starts by initializing an empty feature map and population (Lines~1--2). Then, the \textit{while} loop at Lines~4--9 fills the initial population by randomly generating an individual (Line~5) and executing it to collect its fitness value $f$ (Line~6). 

\begin{algorithm}[t]
	
\DontPrintSemicolon
\footnotesize

\SetKwInOut{Input}{Input}
\SetKwInOut{Output}{Output}
\SetKwRepeat{Repeat}{repeat}{until}

\Input{
	$M$, DNN model under test; \\
	\, $S$, Simulator instance; \\
 	\, $P_s$, Population size; \\
	\, $N$, Number of iterations.
}
\Output{
	$F_m$, feature map.
}

$M$ $\gets$ \textsc{initFeatureMap}() \\
$pop$ $\gets$ $\emptyset$

\HiLi{/* Generate Initial Population */} \\
\While{$i \le P_s$}{
	$t_c$ $\gets$ \textsc{generateIndividual}() \\
	$f$ $\gets$ \textsc{executeIndividual}($t_c$, $M$, $S$) \\
	\textsc{placeIndividualMap}($F_m$, $f$, $t_c$) \\
	$pop$ $\gets$ $pop$ $\cup$ $\{t_c\}$
}

\HiLi{/* Evolve Individuals */} \\
\While{$i \le N$}{
	$t_c$ $\gets$ \textsc{selectIndividual}($pop$) \\
	$\hat{t}_c$ $\gets$ \textsc{mutateIndividual}($t_c$) \\
	$f$ $\gets$ \textsc{executeIndividual}($\hat{t}_c$, $M$, $S$) \\
	\textsc{placeIndividualMap}($F_m$, $f$, $\hat{t}_c$) \\
}

\Return{$F_m$}

\caption{DeepHyperion algorithm}
\label{algorithm:approach:deephyperion}
\end{algorithm}

The assignment to the feature map (Line~7) is done by the procedure \textsc{placeIndividualMap} based on the feature values of the individual $t_c$ (to determine the coordinates of the target cell) and its fitness value.
If the target cell is empty, the individual is placed in the cell. If the cell is non-empty (i.e., another test case was already generated for that cell), a \textit{local competition} based on the value of the fitness takes place. If the fitness of the individual in the cell is greater than the fitness of the candidate individual, the individual in the cell gets replaced with the candidate individual. Otherwise, no replacement is carried out, which also holds if the individual in the cell already has a negative fitness.
The selection function ensures that the search space of the features is explored at large, while the local competition on the individual cells keeps only the lowest performing individuals (i.e., potential misbheaviours) at the end of the generation in order to guide the search towards misbehaviors with unique feature values.

The \textit{while} loop at Lines~11--16 evolves the initial population of individuals. First, an individual is selected (Line~12) and mutated (Line~13), i.e., the control points of the road are changed in order to form a new individual $\hat{t}_c$ with different features. Such individual is then executed (Line~14) and placed in the map (Line~15). The algorithm terminates after a number $N$ of iterations (Line~16).

\autoref{algorithm:approach:deephyperion} returns a feature map with a single individual for each cell, i.e., the one with the lowest fitness (Line~17). In order to further explore the search space, we run DeepHyperion multiple times for each digital sibling to generate multiple feature maps. Then, we combine such maps by considering the \textit{bounds} of each feature map axis in all the runs (i.e., minimum and maximum value), and place each generated individual in the combined map, whose bounds are the lowest (resp. highest) bound values across maps. In this way, there are potentially multiple individuals in each cell, and the value of a cell represents the metric of interest averaged over all individuals in that cell (see $FM_{DS_1}$ and $FM_{DS_2}$ in \autoref{fig:approach:overview}). For instance, considering the failure probability, the value of a cell represents the number of times the model under test failed over the number of all individuals in the cell (a failure occurs when the fitness of an individual is negative).

\subsection{Migration and Union} \label{sec:approach:migration-and-union}

The test generation step produces two feature maps $FM_{DS_1}$ and $FM_{DS_2}$, for \dsone and \dstwo, respectively (in general, N feature maps, i.e., $FM_{DS_1}$, \dots, $FM_{DS_N}$). 
The next step of our approach (i.e., step~\ding{184}, see \autoref{fig:approach:overview}) consists of \textit{migrating} the test cases in $FM_{DS_1}$ to \dstwo (producing $\overline{FM}_{DS_1}$) and viceversa (producing $\overline{FM}_{DS_2}$). In general, migrating the test cases in $FM_{DS_i}$ (with $i = 1, \dots, N$) to DS\textsubscript{j} (with $j \neq i$), would produce $\overline{FM}_{DS_{ij}}$. For instance, if $N = 3$, migrating the test cases in $FM_{DS_2}$ to the other siblings, would produce $\overline{FM}_{DS_{21}}$ when migrating to DS\textsubscript{1}, and $\overline{FM}_{DS_{23}}$ when migrating to DS\textsubscript{3}. Such operation consists of instantiating the abstract (control point based) road representation of the test case being migrated, such that it respects the dimensionality constraints, and it can be supplied as input to the target simulator.

After migration, for both \dsone and \dstwo (in general, DS\textsubscript{1}, \dots, DS\textsubscript{N}), we consider the \textit{union} of their maps. We consider the bounds of each feature in the two maps, and we place the respective test cases in a new unified map according to their coordinates, producing the map $FM_{U_1}$ for \dsone (i.e., $FM_{DS_1} + \overline{FM}_{DS_2}$) and the map $FM_{U_2}$ for \dstwo (i.e., $FM_{DS_2} + \overline{FM}_{DS_1}$). In general, $FM_{U_i} = FM_{DS_i} + \sum_{j \neq i} \overline{FM}_{DS_{ji}}$. For instance, if $N = 3$, $FM_{U_2} = FM_{DS_2} + (\overline{FM}_{DS_{12}} + \overline{FM}_{DS_{32}})$. Hence, the two maps, or $N$ maps in general, contain the same tests that fill the same cells at the same coordinates.

The value of each cell in the union maps $FM_{U_1}$, $FM_{U_2}$ is recomputed from the individuals assigned to them. For the failure probability, if a given cell in $FM_{DS_1}$ has $n_1/N_1$ failing individuals, while the corresponding cell in $\overline{FM}_{DS_2}$ has $n_2/N_2$ failing individuals, the failure probability value of the cell in the union map $FM_{U_1}$ will be $(n_1 + n_2) / (N_1 + N_2)$. In general, for a given cell in $FM_{U_i}$, the failure probability is computed as $(n_1 + \dots + n_i + \dots n_N) / (N_1 + \dots + N_i + \dots + N_N)$. When a quality of driving metric is computed instead of a failure probability, the union map contains the average of the respective quality of driving metrics: $qm = (qm_1 + qm_2) / 2$, where $qm_1$, $qm_2$ are the quality of driving metrics found in the same cell in the two feature maps being united ($FM_{DS_1}, \overline{FM}_{DS_2}$, or $FM_{S_2}, \overline{FM}_{S_1}$), while $qm$ is the resulting quality of driving metric, in the union map ($FM_{U_1}$ or $FM_{U_2}$). In general, for a given cell in $FM_{U_i}$, the quality metric is computed as $(qm_1 + \dots + qm_i + \dots qm_N) / N$.

\subsection{Merge}

The final step of the approach (i.e., step~\ding{185} in \autoref{fig:approach:overview}) requires to \textit{merge} the two union maps $FM_{U_1}$ and $FM_{U_2}$ into $FM_{DSS}$ (in general, $N$ union maps $FM_{U_1}$, \dots, $FM_{U_N}$). The objective of the merge operation is to combine the testing output of the two digital siblings.
Since we aim to use the digital siblings to approximate the behavior of $M$ on \hdt and predict its failures, the merge operator privileges \textit{agreements} between the maps of the two digital siblings, i.e., only cells in the maps that have a hot color (e.g., a high failure probability) will produce a hot color in the merged cell. Indeed, such tests are likely to represent simulator-independent misbehaviors of the model under test, which are critical for the safety of the system. 
Specifically, if the failure probability of $FM_{U_1}$ is $fp_1 = n_1/N_1$ and that of $FM_{U_2}$ is $fp_2 = n_2/N_2$, in the merged map the failure probability will be the product, $fp = fp_1 \times fp_2$ (in general, the failure probability of a given cell in \dss would be $fp = fp_1 \times \dots \times fp_i \times \dots \times fp_N$). When a quality of driving (resp. lack of quality of driving) metric is computed instead of a failure probability, the merged map will conservatively contain the maximum (resp. minimum) of the respective quality of driving metrics. In particular, $qm = \max\{qm_1, qm_2\}$ (resp. $qm = \min\{qm_1, qm_2\}$), where $qm_1$, $qm_2$ are the quality of driving metrics found in the same cell in $FM_{U_1}$ and $FM_{U_2}$ respectively, while $qm$ is the resulting quality of driving metric in the merged map. In general, the quality metric of a given cell in \dss would be $qm = \max\{qm_1, \dots, qm_i, \dots, qm_N\}$, and the lack of quality of driving of a given cell would be $qm = \min\{qm_1, \dots, qm_i, \dots, qm_N\}$.
By giving  priority to failures (resp. quality of driving degradations) that occur in both siblings and are hence very likely to be relevant for the target platform, this choice better accommodates the limited testing budget available for production/field testing~\cite{waymo,borg,comprehensive-sfc-test,outsourcing-av-development,stocco-mind}.

\subsection{Evaluation Scenario}

While our approach assumes that \hdt is not available in practice, to evaluate whether the \dss can approximate the behavior of $M$ and predict its failures when executed on \hdt, we migrate all the tests in the digital siblings feature map (i.e., $FM_{DSS}$) to an actual \hdt, which is used  to obtain the ground truth map $FM_{\hdt}$ (see ``Evaluation Scenario'' in \autoref{fig:approach:overview} (bottom)).
The two maps being compared contain the same tests in the same cells, but the values of the cells might differ, depending on the behavior of $M$ in the different simulators. Thus, we analyze and compare the two feature maps $FM_{DSS}$ and $FM_{\hdt}$, to assess the capability of \dss at predicting the failures of the model when executed on \hdt.

\section{Case Study}\label{sec:study}

The goal of the empirical study is to evaluate whether two digital siblings (\dss) can better approximate the \textit{behavior} of a driving model and predict its failures on a digital twin (\hdt), w.r.t. using only one general-purpose simulator (\gps). We rely on \hdt only to evaluate the benefits of our multi-simulator approach, as a proxy for the behaviors of the AV in the real world, since \hdt is often unavailable in practice.
In our empirical study, we focus on testing a lane-keeping DNN model by generating road scenarios.
To this aim, we consider the following research questions: 

\noindent
\head{RQ\textsubscript{1} (Offline Evaluation)}
\textit{How do the offline prediction errors by the \dss compare to those of the \hdt?}

We first test our hypothesis at the model-level. 
For all simulators, we compute the errors between the model predictions and each autopilot ground truth labels on a stationary driving images dataset. We compare the error distributions of each individual simulator with the \hdt, as well as their combination as digital siblings. 

With RQ\textsubscript{1} we aim to assess whether a correlation between the offline predictions exists at the model-level, which can be useful for developers to gain trust about their DNN model prediction accuracy, prior to running system-level tests.

\noindent
\head{RQ\textsubscript{2} (Failure Probability)}
\textit{How does the failure probability of the \dss compare to that of the \hdt?}

In RQ\textsubscript{2} we test the model at the system-level, specifically the hypothesis that combining the failure probabilities of the two digital siblings provides a better predictor of the ground truth failure probability of the model executed on \hdt w.r.t. using a single simulator. A positive answer to RQ\textsubscript{2} would imply that a multi-simulator approach can predict, and possibly anticipate, the failures of the DNN-based lane-keeping model on \hdt, which are expected to be accurate proxies of the AV real-world failures.

\noindent
\head{RQ\textsubscript{3} (Quality of Driving)}
\textit{How does the quality of driving of the \dss compare to the failure probability of the \hdt?}

By considering only the failure probability, we might overlook the correlation between real failures on \hdt and near-failures on \dss---test cases in which the model exhibits a degraded driving quality without necessarily going off-road. Thus, with RQ\textsubscript{3}, we also assess whether finer-grained driving quality metrics can predict the ground truth failure probability of the lane-keeping model on \hdt.

\subsection{Test Object and Simulators}

\subsubsection{Study Object}
We considered three self-driving architectures, i.e., \davetwo~\cite{nvidia-dave2}, \chauffeur~\cite{chauffeur} and \epoch~\cite{epoch}. Such architectures were used in previous studies on AV testing in the literature~\cite{stocco-mind,2020-Stocco-ICSE,survey-lei-ma,2021-Jahangirova-ICST,2022-Stocco-ASE,2020-Stocco-GAUSS,2021-Stocco-JSEP,deephyperion,sbst2021,sbst2022,sbst2023}, and the respective models feature different number of parameters. The \davetwo model has 2.8$M$ parameters, \chauffeur has 100$k$ parameters while \epoch has 26$M$ parameters (we used a reduced version of the \epoch model to reduce training and inference time~\cite{stocco-mind}).

Architecturally, \davetwo consists of five convolutional layers, followed by three fully-connected layers~\cite{nvidia-dave2}. \chauffeur has six convolutional layers each followed by a dropout and a max pooling layer (except the last one)~\cite{chauffeur}. \epoch has three convolutional layers and one fully-connected layer, which makes up for most of the parameter count of the model~\cite{epoch}.

\subsubsection{Digital Siblings (DSS)}
We implemented and investigated the effectiveness of \dss using the simulators BeamNG~\cite{beamng} and Udacity~\cite{udacity-simulator}. 
We chose them as digital siblings because: 
(1)~they support training and testing of a DNN that performs lane-keeping, including \davetwo, \chauffeur and \epoch;
(2)~they are often used as simulator platforms for AV testing, as highlighted by a recent survey on autonomous driving testing~\cite{survey-lei-ma};
(3)~they are potentially complementary because they are developed with different technologies/game engines, and they are characterized by different physics implementations (e.g., rigid vs soft-body dynamics);   
(4)~they are publicly available under open-source or academic-oriented licenses, hence customizable. 

BeamNG~\cite{beamng} is a framework specialized in autonomous driving developed by BeamNG GmbH. The framework is released under an academic-oriented license, and it has been downloaded 5.5$k$ times as of January 2023. From a technical standpoint, BeamNG features a \textit{soft-body dynamics} simulation based on a spring-mass model. Such a model is composed of nodes (mass points) that are connected by beams (springs), i.e., weightless elements that allow accurate vehicle deformation and other aerodynamic properties~\cite{gambi-beamng}.

Udacity~\cite{udacity-simulator} is developed with Unity 3D~\cite{unity}, a popular cross-platform game engine. The project has been publicly released in 2016 by the for-profit educational organization Udacity, to allow people from all over the world to access some of their technology and to contribute to an open-source self-driving car project. As of January 2023, the simulator has 3.7$k$ stars on GitHub. From a technical standpoint, Udacity is based on the Nvidia PhysX engine~\cite{PhysX}, featuring discrete and continuous collision detection, ray-casting, and \textit{rigid-body dynamics} simulation. 
 
\subsubsection{Digital Twin (\hdt)}
We use the Donkey Car\textsuperscript{\texttrademark} open-source framework~\cite{donkeycar} as digital twin for our study. This platform has been used for AV testing research with physical self-driving cars in physical environments~\cite{stocco-mind,viitala,9412011}. The framework includes open hardware to build 1:16 scale radio-controlled cars with self-driving capabilities, a Python framework for training and testing DNN models with lane-keeping functionalities using supervised or reinforcement learning, and a simulator in which the real-world \dk is faithfully modeled. This was assessed by a recent work~\cite{stocco-mind} reporting that, for three lane-keeping models, the steering angle distribution of the AV model driving in the real-world environment is statistically indistinguishable from the steering angle distribution of the AV model driving in the digital twin.

In the rest of the section, we refer to BeamNG as \dsone, Udacity as \dstwo, the combined digital siblings as \dss, and DonkeyCar as \hdt. 

\subsection{Procedure} \label{sec:study:procedure}

\subsubsection{CycleGAN Models}

\head{Data Collection} We collected 15\textit{k} simulated images, 5\textit{k} for \dsone and  \dstwo by running the autopilots on a set of randomly generated roads.
Moreover, we collected 5\textit{k} real-world images~\cite{stocco-mind} by manually driving the physical twin of the \hdt on a physical road track in our lab. 

\head{Training} We trained three CycleGAN models, one for each simulator, with the obtained training sets (5\textit{k} virtual images and 5\textit{k} real-world images).  
Each model was trained for 60 epochs using the default hyper-parameters of the original paper~\cite{cyclegan}. We saved a checkpoint model every 5 epochs, and we ultimately chose the one that achieved the best neural translations (in terms of visual quality) using a test set of $\approx$8\textit{k} simulated images for each simulator, representing a test road driven from beginning to the end~\cite{stocco-mind}.
While a quantitative assessment of the output of CycleGAN is still a major challenge~\cite{pros-and-cons-gans,2024-Lambertenghi-ICST} and out of the scope of this paper, 
the driving capability of the lane-keeping model, as the experimental evaluation shows, represents an implicit validation of the CycleGAN model's ability to retain all essential features needed for an accurate steering angle prediction. 

\subsubsection{Driving Models}

\head{Data Collection} For all simulators (i.e., \dsone, \dstwo and \hdt), we collected a training set by running the autopilots on a set of randomly generated roads (this set is different from the one used to train the CycleGAN).
To ensure having non-trivial driving scenarios and appropriate labels for challenging curves, the maximum angle of a curve was set to be less than or equal to $270^\circ$. 
In particular, for our training set, we generated 25 roads with 8 control points~\cite{deephyperion}.
To collect a balanced dataset where left and right curves are equally represented, each road was driven by the autopilot in both directions, i.e., from the start point to the end point and from the end point to the start point. The autopilot drove successfully the totality of the roads on all simulators; our training set comprises $\approx$70\textit{k} images, equally distributed across the simulators. 

\head{Training} 
For each self-driving architecture we trained two models, one by using the plain simulated images (\msim) and the other by translating the images of each simulator into \textit{pseudo}-real images (\mreal) using the respective CycleGAN generator.

We followed the guidelines by Bojarski et al.~\cite{nvidia-dave2} to train AV autopilots.
We used custom hyperparameters for each self-driving architecture, and the Adam optimizer~\cite{kingma2014adam} to minimize the mean squared error (MSE) between the predicted steering angles and the ground truth value. For all models, we set a learning rate of $10^{-4}$ and a batch size of 128. We used 50 epochs for \davetwo and \chauffeur (only for the \mreal model) and 500 epochs for \epoch and the \msim model of \chauffeur. We used an early stopping of 10 epochs for the models where the number of training epochs was 50 and an early stopping of 20 epochs otherwise.

\begin{table}[ht]
	\centering
	
	\caption{Offline and online performance on the test set of the lane-keeping models on \hdt.}
	\label{table:study:models}
	\setlength{\tabcolsep}{4pt}
	\renewcommand{\arraystretch}{1.5}
	\color{myblue}
	\centering

		\begin{tabular}{lcccc}
			\toprule

			\multicolumn{1}{l}{} 
			& \multicolumn{2}{c}{\msim} 
			& \multicolumn{2}{c}{\mreal} \\
			
			\cmidrule(r){2-3} 
			\cmidrule(r){4-5}
			
			\multicolumn{1}{l}{} 
			& MSE
			& Success Rate
			& MSE
			& Success Rate \\
			
			\midrule
			
			\davetwo~\cite{nvidia-dave2} & 0.08 & 0.84 & 0.07 & 0.96 \\
			\chauffeur~\cite{chauffeur} & 0.07 & 0.72 & 0.07 & 0.92 \\
			\epoch~\cite{epoch} & 0.09 & 0.52 & 0.07 & 0.96 \\
			
			\midrule
			
			Avg & 0.08 & 0.69 & 0.07 & 0.95 \\
			
			\bottomrule
			
		\end{tabular}
		
	\end{table}

We evaluated the performance of the trained lane-keeping models on \hdt, as it is the target simulator we want to approximate using the digital siblings. We collected a labeled dataset by running the autopilot on \hdt on 25 randomly generated roads each with 8 control points and a maximum angle of  $270^\circ$, i.e., the same road parameters as the training set. We computed the mean squared error (MSE) between the steering angle prediction of the model on each image and the steering angle of the autopilot. \autoref{table:study:models} shows the MSE of all models on the first and third columns; on average, the MSE is low for both the models trained using simulated images (i.e., \msim), and the models trained using real images (i.e., \mreal). We also measured the success rate of each model by driving it on the 25 randomly generated roads, and counting the number of times the model was able to arrive at the end of the road without going out of bound. Overall, each model is able to successfully complete the majority of the generated roads. Most notably, \mreal models are able to complete more than 90\% of the test set roads.

\subsubsection{Offline Evaluation} 

We collected a labeled dataset for offline evaluation by generating 20 roads (i.e., 10 roads driven in both directions) with the same parameters as the training set. The images collected for the \textit{offline} evaluation dataset amount to $\approx$26\textit{k}, considering all simulators.

\subsubsection{Test Generation} 

After training \msim and \mreal for each self-driving architecture, we executed DeepHyperion \textit{twice} to generate tests using the two digital siblings \dsone and \dstwo. 
We chose a population size of 20 individuals and a number of search iterations respectively equal to 150 for \msim and 100 for \mreal, as we observed from preliminary experiments that this choice of hyperparameters allows an extensive coverage of the feature maps. For both \msim and \mreal and each digital sibling in each self-driving architecture, we repeated test generation five times to diversify the exploration of the search space and to collect multiple test cases for each cell in the feature maps. Overall, across all runs and driving models, DeepHyperion generated 10,260 tests for both siblings. 

Concerning the simulations, for all simulators, we set the maximum speed for the vehicle to 30 km/h~\cite{deephyperion}.
When testing \mreal in a given simulator, we engineered the testing pipeline to load the appropriate sim2real CycleGAN generator to translate the simulated image generated by BeamNG/Udacity into pseudo-real images \textit{in real-time during driving}. 
For each executed test case, we collected the lateral position of the vehicle for each simulation step as well as its lateral distance. The former determines the quality of driving of the model~\cite{2021-Jahangirova-ICST}, while the latter is the fitness of the test case.

\subsubsection{Migration and Union}

For the initial ($FM_{DS_1}$, $FM_{DS_2}$) and for the union ($FM_{U_1}$, $FM_{U_2}$) feature maps, we compute the failure probability as the number of tests with a negative fitness divided by the total number of tests in the respective cell. 
To evaluate the quality of driving, we adopted the maximum lateral position (i.e., the distance between the center of the vehicle and the center of the lane~\cite{2020-Stocco-GAUSS}) experienced during the test case execution. Previous work showed that such metric is effective at characterizing the degradation in the quality of autonomous driving~\cite{2021-Jahangirova-ICST}, since the lower the value of such metric, the higher is the quality of driving (thus, it actually measures \textit{lack} of quality of driving).
When considering the quality of driving, the value of each cell in a feature map represents the average of the maximum lateral positions of each test case in that cell. Furthermore, we normalized the maximum lateral position values in the interval $[0,1]$ before taking the union.

\subsubsection{Merge}

Merging the maps of the two digital siblings requires a different treatment for failure probability and quality of driving. Regarding the failure probability, the merge operator that ensures a conservative aggregation of two values is the \textit{product}. 
Regarding the lack of quality of driving, the conservative merge operator is the \textit{minimum}, since the quantities to merge are not probabilities. 
In fact, by taking the minimum we get a high lack of driving quality only when both simulators exhibit high values for such a metric.

\subsection{Metrics}

\subsubsection{RQ\textsubscript{1} (Offline Evaluation)}
We computed the prediction errors given by the difference between the predictions of the model (\mreal) on images of the offline evaluation dataset (see \autoref{sec:study:procedure}), and the corresponding ground truth labels given by the autopilot. We binned the prediction errors of the model on each simulator and built the respective \textit{probability density} (i.e., the number of errors in each bin is divided by the total number of prediction errors) such that different distributions could be compared.

Then, we computed the \textit{distance} between each digital sibling distribution, as well as their combination, and the \hdt using the \textit{Wasserstein} distance~\cite{wgan} (also known as the \textit{earth mover's distance}). 
Given two one-dimensional distributions $A$ and $B$, the Wasserstein distance $W(A,B)$ is defined by the following formula~\cite{wasserstein}:

\begin{equation} \label{eq:study:wasserstein}
	W(A,B) = \int_{\mathbb{R}} |CDF_A(x) - CDF_B(x)| dx
\end{equation}
\noindent
where $CDF$ is the \textit{cumulative distribution function} of a distribution. In other words, the Wasserstein distance between two distributions is defined as the difference between the area formed by their cumulative distribution functions.

We assess whether the difference between two distributions is statistically significant using the Wilcoxon test~\cite{non-parametric-book} applied to the density functions of the two error distributions to compute the $p$-value (with threshold $\alpha \le 0.05$). We also perform power analysis (with statistical power $\beta \ge 0.8$) on the prediction errors to check whether a non-significant $p$-value is due to a low data sample size or to the difference being statistically insignificant.

\subsubsection{RQ\textsubscript{2} (Failure Probability) and RQ\textsubscript{3} (Quality of Driving)}
For RQ\textsubscript{2}, we computed the pairwise \textit{Pearson correlation} between maps along with the corresponding $p$-value. In particular, correlations are computed between each union feature map of each digital sibling ($FM_{U_1}$, $FM_{U_2}$) and the feature map of the \hdt ($FM_{\hdt}$), and between $FM_{DSS}$ and $FM_{\hdt}$. 
For RQ\textsubscript{3}, the setting is equivalent to that of the failure probability but considering quality of driving maps, comparing \dsone, \dstwo and \dss against the ground truth \hdt.

To evaluate the capabilities of the digital siblings (individually or jointly) to predict failures on \hdt, we computed the area under the curve Precision-Recall (AUC-PRC) at increasing thresholds, for both RQ\textsubscript{2} and RQ\textsubscript{3}. 
This requires the discretization of failure probabilities into binary values (failure vs non-failure) for the ground truth (i.e., \hdt): we consider a cell in the \hdt feature map to be a failure cell if the associated failure probability is $> 0.0$. 
AUC-PRC is more informative than the AUC-ROC metric (i.e., the area under of the curve of the Receiver Operating Characteristics) when dealing with imbalanced~\cite{prc-justification} datasets, which is the case of our study (the number of failures in the feature maps is lower than the number of non-failures with an average 10 to 20\% ratio).

\begin{table}[ht]
	\centering
	\begin{threeparttable}[b]
	
	\captionsetup{labelfont={color=myblue}}
	\caption{\textcolor{myblue}{Results for RQ\textsubscript{1}. Bold-faced values indicate the best approach.}}
	\label{table:study:rq1}
	\setlength{\tabcolsep}{11pt}
	\renewcommand{\arraystretch}{1.5}
	\color{myblue}
	\centering

	\begin{tabular}{lrllll}
		\toprule
		
		\multicolumn{1}{r}{\textbf{}} &  & \multicolumn{4}{c}{\sc Offline Evaluation (RQ\textsubscript{1})} \\
		
		\cmidrule(r){3-6} 

		\multicolumn{1}{l}{} &  
		& \multicolumn{2}{c}{\msim} 
		& \multicolumn{2}{c}{\mreal} \\
		
		\cmidrule(r){3-4} 
		\cmidrule(r){5-6}

		\multicolumn{1}{l}{} &  
		& distance 
		& p-value
		& distance 
		& p-value \\
		
		\midrule
		
		\multirow{3}{*}{\davetwo~\cite{nvidia-dave2}} & DS\textsubscript{1} vs \hdt & 0.04669 & 0.101 & 0.03250 & 0.011 \\
		& DS\textsubscript{2} vs \hdt & 0.02648 & 0.020 & 0.02187 & 0.078 \\
		& DSS vs \hdt & \textbf{0.03776} & 0.053\tnote{$\dagger$} & \textbf{0.00951} & 0.088\tnote{$\dagger$} \\
		
		\multicolumn{1}{l}{} &  & \multicolumn{1}{l}{} & \multicolumn{1}{l}{} & \multicolumn{1}{l}{} & \multicolumn{1}{l}{} \\
		
		\multirow{3}{*}{\chauffeur~\cite{chauffeur}} & DS\textsubscript{1} vs \hdt & 0.03989 & 0.023 & 0.04625 & 0.011 \\
		& DS\textsubscript{2} vs \hdt & 0.02641 & 0.047 & 0.02145 & 0.078\tnote{$\dagger$} \\
		& DSS vs \hdt & \textbf{0.01208} & 0.394\tnote{$\dagger$} & \textbf{0.01843} & 0.334\tnote{$\dagger$} \\
		
		\multicolumn{1}{l}{} &  & \multicolumn{1}{l}{} & \multicolumn{1}{l}{} & \multicolumn{1}{l}{} & \multicolumn{1}{l}{} \\
		
		\multirow{3}{*}{\epoch~\cite{epoch}} & DS\textsubscript{1} vs \hdt & 0.06030 & 0.011 & 0.03374 & 0.016 \\
		& DS\textsubscript{2} vs \hdt & \textbf{0.01634} & 0.078\tnote{$\dagger$} & 0.02318 & 0.078\tnote{$\dagger$} \\
		& DSS vs \hdt & 0.02726 & 0.053\tnote{$\dagger$} & \textbf{0.00989} & 0.256\tnote{$\dagger$} \\
		
		\bottomrule
		
	\end{tabular}
	\begin{tablenotes}[para]
		\item[$\dagger$] \textit{power $>$ 0.8}
	\end{tablenotes}
	\end{threeparttable}

\end{table}

\subsection{Results}

\subsubsection{Offline Evaluation (RQ\textsubscript{1})} 
\autoref{table:study:rq1} reports the results for our first research question. The first column shows the simulators being compared. Columns 2--5 report the Wasserstein distance between the prediction error densities of the corresponding simulators, and the $p$-value concerning the statistical significance of the differences between the two densities, for \msim and \mreal. 

For \msim (Columns 3--4), our results show that, for \davetwo, the distance between the steering angle errors obtained for the combined digital siblings \dss and the errors obtained for \hdt is lower than the distance of \dsone (0.03776 vs 0.046) and higher than the distance of \dstwo (0.02648). The distribution of the steering angle errors of \dstwo is statistically different from the errors of \hdt (i.e., $p$-value 0.02 $< 0.05$), while the distribution of the steering angle errors of \dss is statistically indistinguishable from the errors of \hdt (i.e., $p$-value 0.053 $> 0.05$ and power $> 0.8$). This behavior is also consistent for \epoch, with the exception that the distribution of the prediction errors for \dstwo is statistically indistinguishable from that of \hdt. However, the distance between \dss and \hdt is lower than the distance of \dsone from \hdt, with a statistically indistinguishable distribution of prediction errors w.r.t. \hdt. For \chauffeur, the combined digital siblings \dss have the only distribution of errors that is equivalent to that of \hdt, and its distance to it is the lowest considering the individual digital siblings.

Regarding \mreal (Columns 5--6), our results show that, for \davetwo, the distance between the steering angle errors obtained for the combined digital siblings \dss and the errors obtained for \hdt is \textit{2.8 times lower} than the distance of each simulator taken individually (as a percentage, the distance of \dss is respectively 70\% and 56\% smaller than the distance of the two individual siblings, DS\textsubscript{1}, DS\textsubscript{2}). 
The statistical test confirms that the error distributions of \dss and \hdt are statistically indistinguishable ($p$-value $>$ 0.05 and power $>$ 0.8), which is not the case for the error distributions of DS\textsubscript{1} ($p$-value $<$ 0.05). Likewise, for all the other self-driving architectures, the digital siblings \dss have the lowest distance to \hdt w.r.t. the individual siblings and their distribution is always statistically indistinguishable from that of \hdt.

\begin{figure*}[t]
\centering

\includegraphics[trim=1.8cm 25cm 54cm 0cm, clip=true, scale=0.135]
{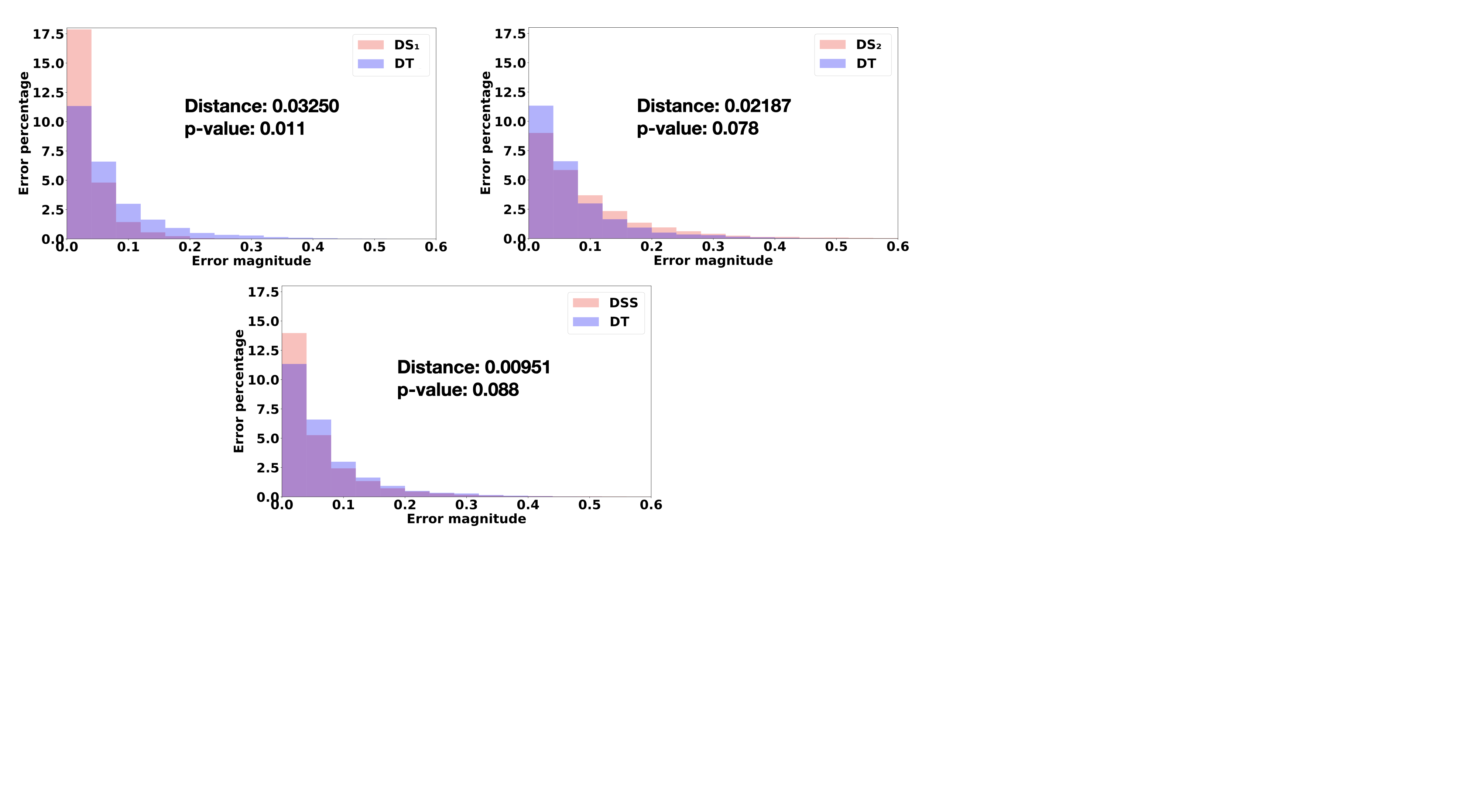}

\caption{Distributions of prediction errors of \davetwo \mreal in the two digital siblings, i.e., DS\textsubscript{1} and DS\textsubscript{2}, their combination (\dss) and \hdt. Best viewed in color.} 
\label{fig:study:offline-evaluation} 
\end{figure*}

\autoref{fig:study:offline-evaluation} offers a visual explanation of these scores for the \davetwo model.\footnote{We report the plots for the other lane-keeping models in our replication package~\cite{tool}.}
The subplots compare the steering angle error distributions, respectively, of \dsone, \dstwo and \dss (shown in light red) with that of \hdt (shown in light blue). The $x$-axis of each subplot represents the magnitude of the prediction errors of the model \mreal w.r.t. the predictions of the autopilot, while the $y$-axis indicates their percentage for each bin.

From the plots we can see that, overall, at the model-level, \mreal makes prediction errors with small magnitudes on DS\textsubscript{1}, DS\textsubscript{2} and \dss (i.e., most of the errors are between 0.0 and 0.3).
On the digital sibling DS\textsubscript{1} (i.e., BeamNG), \mreal has a high agreement with the autopilot, as most errors have a low magnitude. It has numerous small errors ($< 0.2$), while it has only a negligible portion of the distribution being above 0.2. 
The agreement with \hdt is low as \mreal \textit{under-approximates} the true error distribution on \hdt: \mreal on \hdt has fewer errors with low magnitude and has a longer tail of errors greater than 0.2 (even greater than 0.3 in some cases). 
Differently, on the digital sibling DS\textsubscript{2} (i.e., Udacity), the error distribution has a longer tail than that on \hdt.
Indeed, \mreal executed on DS\textsubscript{2} \textit{over-approximates} the errors it would have on \hdt, as the errors observed on DS\textsubscript{2} have higher magnitude than those observed on \hdt. 

The error distribution of the model on \dss shows why it is appropriate to combine the outcome of two simulators. 
At the model-level, \dss better approximates the true error distribution of the model on \hdt, by providing an intermediate error between DS\textsubscript{1} and DS\textsubscript{2} for both \msim and \mreal.

\begin{tcolorbox}[colback=gray!15!white,colframe=black]
	\textbf{RQ\textsubscript{1}}: Overall, at the model-level, the digital siblings produce a steering angle error distribution that is statistically indistinguishable from the true steering angle error distribution of the model on the digital twin. Considering all the models, in 5 out of 6 cases, the digital siblings are better at approximating the distribution of prediction errors of the digital twin than each individual sibling.
\end{tcolorbox}

\subsubsection{Failure Probability (RQ\textsubscript{2})}
\autoref{table:study:rq2} shows the Pearson correlation (\textit{r}), the $p$-value, and the AUC-PRC for the comparison between DS\textsubscript{1}, DS\textsubscript{2}, \dss and \hdt, respectively. The analysis is reported separately for \msim (Columns 3--5) and \mreal (Columns 6--8). 

\begin{table}[ht]
	
	\captionsetup{labelfont={color=myblue}}
	\caption{\textcolor{myblue}{Results for RQ\textsubscript{2}. Bold-faced values indicate the best approach.}}
	\label{table:study:rq2}
	\setlength{\tabcolsep}{4.5pt}
	\renewcommand{\arraystretch}{1.5}
	\color{myblue}
	\centering

	\begin{tabular}{lrllllll}
		\toprule
		
		\multicolumn{1}{r}{\textbf{}} & 
		& \multicolumn{6}{c}{\sc Failure Probability (RQ\textsubscript{2})} \\
		
		\cmidrule(r){3-8} 
		
		\multicolumn{1}{l}{} &  
		& \multicolumn{3}{c}{\msim} 
		& \multicolumn{3}{c}{\mreal} \\
		
		\cmidrule(r){3-5}
		\cmidrule(r){6-8}
		
		\multicolumn{1}{l}{} &  
		& \textit{r} 
		& p-value 
		& AUC-PRC 
		
		& \textit{r} 
		& p-value 
		& AUC-PRC \\
		
		\midrule
		
		\multirow{3}{*}{\davetwo~\cite{nvidia-dave2}} & DS\textsubscript{1} vs \hdt & 0.650 & $10^{-11}$ & 0.654 & 0.391 & $10^{-4}$ & \textbf{0.403} \\
		& DS\textsubscript{2} vs \hdt & 0.583 & $10^{-8}$ & 0.512 & 0.377 & $10^{-4}$ & 0.306 \\
		& DSS vs \hdt & \textbf{0.710} & $10^{-13}$ & \textbf{0.684} & \textbf{0.457} & $10^{-5}$ & 0.398 \\

		\multicolumn{1}{l}{} &  & \multicolumn{1}{l}{} & \multicolumn{1}{l}{} & \multicolumn{1}{l}{} & \multicolumn{1}{l}{} & \multicolumn{1}{l}{} & \multicolumn{1}{l}{} \\
		
		\multirow{3}{*}{\chauffeur~\cite{chauffeur}} & DS\textsubscript{1} vs \hdt & \textbf{0.733} & $10^{-16}$ & \textbf{0.774} & 0.417 & $10^{-4}$ & 0.481 \\
		& DS\textsubscript{2} vs \hdt & 0.588 & $10^{-10}$ & 0.715 & 0.337 & $10^{-3}$ & 0.300 \\
		& DSS vs \hdt & 0.700 & $10^{-14}$ & 0.742 & \textbf{0.422} & $10^{-4}$ & \textbf{0.496} \\
		
		\multicolumn{1}{l}{} &  & \multicolumn{1}{l}{} & \multicolumn{1}{l}{} & \multicolumn{1}{l}{} & \multicolumn{1}{l}{} & \multicolumn{1}{l}{} & \multicolumn{1}{l}{} \\
		
		\multirow{3}{*}{\epoch~\cite{epoch}} & DS\textsubscript{1} vs \hdt & 0.561 & $10^{-8}$ & 0.599 & 0.469 & $10^{-5}$ & 0.586 \\
		& DS\textsubscript{2} vs \hdt & 0.428 & $10^{-5}$ & 0.604 & \textbf{0.521} & $10^{-7}$ & 0.565 \\
		& DSS vs \hdt & \textbf{0.571} & $10^{-8}$ & \textbf{0.622} & 0.450 & $10^{-5}$ & \textbf{0.641} \\
		
		\bottomrule
		
	\end{tabular}

\end{table}

Concerning \msim---i.e., the model driving with simulated driving scenes---
the failure probabilities for \davetwo have a high positive correlation with the true failure probability of \hdt ((Column~3). All such correlations are statistically significant for \dss, as well as for each individual sibling DS\textsubscript{1} and DS\textsubscript{2} ($p$-values $<$ 0.05, see Column~4). Likewise, the correlations are high and statistically significant for the other lane-keeping models (\epoch features slightly lower correlations).
\begin{figure*}[b]
\centering

\includegraphics[trim=1cm 0cm 46cm 0cm, clip=true, scale=0.12]
{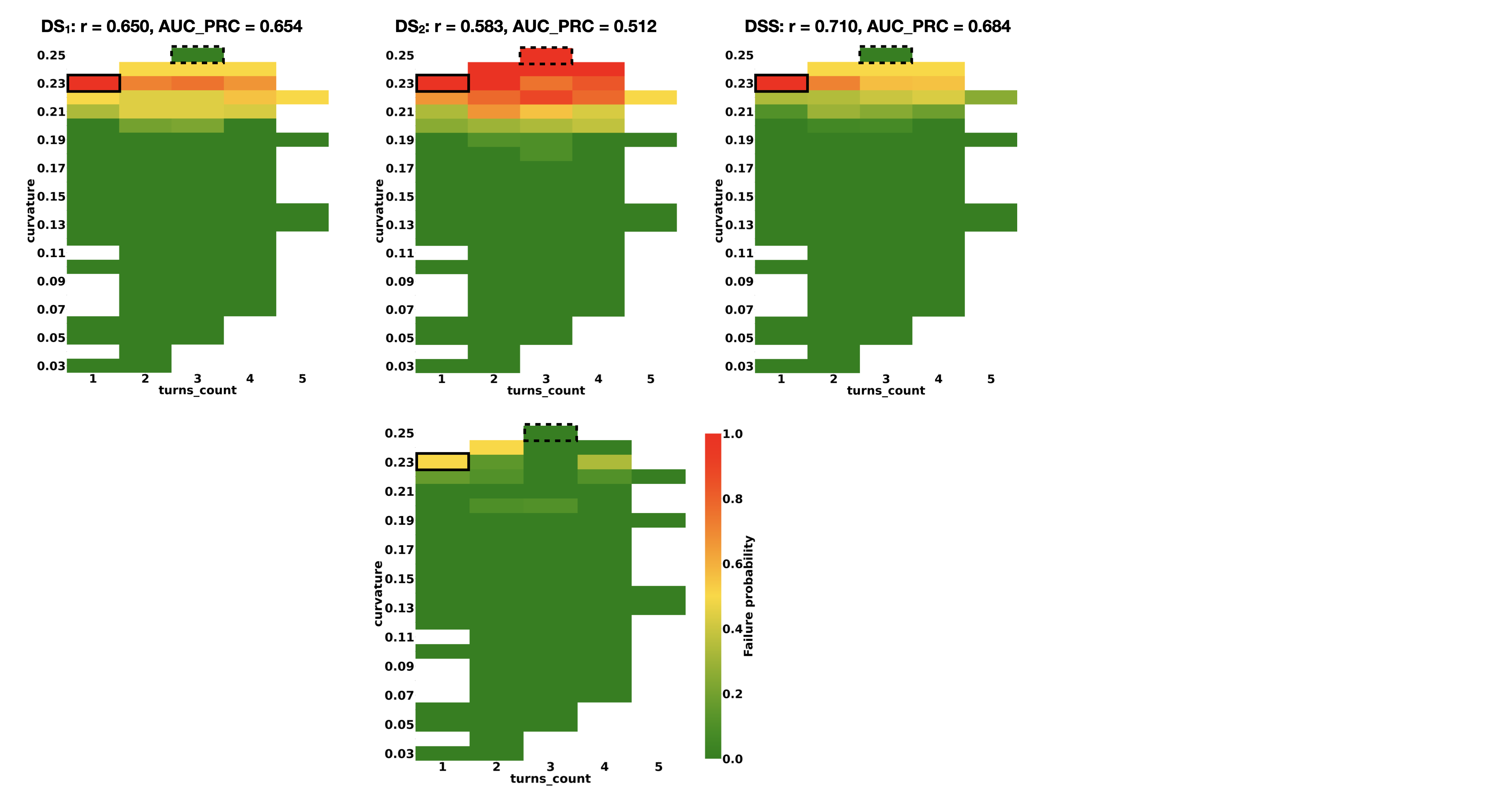}

\caption{Feature maps representing the failure probability of \davetwo \msim on the two digital siblings, DS\textsubscript{1} and DS\textsubscript{2}, their combination (\dss) and on \hdt. Solid line cells represent a true failure predicted by \dss while dashed line cells represent a false positive of \dstwo. Best viewed in color.} 
\label{fig:study:fp-sim} 
\end{figure*}

However, for \davetwo the correlation of \dss is 9\% higher than the best individual correlation (i.e., DS\textsubscript{1}) and 21\% higher than the worst individual correlation (i.e., DS\textsubscript{2}). 
In terms of failure prediction, \dss have the highest AUC-PRC value, 4\% higher than DS\textsubscript{1} and 33\% higher than DS\textsubscript{2}. 

This also happens with \epoch, where the correlation of \dss is slightly higher than that of the best sibling DS\textsubscript{1} (i.e., 0.571 vs 0.561) and 33\% higher than that of the worst sibling DS\textsubscript{2}. Regarding failure prediction on \hdt, \dss are 3\% better than the best sibling. In the case of \chauffeur, DS\textsubscript{1} has the best results both in terms of correlation and failure prediction. However, \dss are better than the worst of the two siblings DS\textsubscript{2} both in terms of correlation and failure prediction.

\autoref{fig:study:fp-sim} shows the feature maps related to \msim of \davetwo.\footnote{We report the plots for the other lane-keeping models in our replication package~\cite{tool}.} The first three feature maps represent the failure probability of DS\textsubscript{1}, DS\textsubscript{2} and \dss, respectively. The last feature map represents the ground truth failure probability of \hdt. 
The color of each cell ranges from green (i.e., non-failure, or failure probability = 0) to red (i.e., failure probability = 1). 
Let us analyze a \textit{false positive} case. The test cases at coordinates (3, 0.25), whose corresponding cells are highlighted with a dashed line, represent road tracks having three curves and a maximum curvature of 0.25. In \hdt, this cell is green, i.e., all test cases for \msim driving on \hdt succeed. On the other hand, \msim has contrasting behaviors when the same test cases are executed on DS\textsubscript{1} or DS\textsubscript{2}. These test cases did not exhibit any failure in DS\textsubscript{1}, whereas they did trigger failures in DS\textsubscript{2}. This disagreement is canceled out when combining the two digital siblings with the product operator and the cell is green in the \dss map.  As such, digital siblings are conservative w.r.t. failures, as a failure is reported only when both digital siblings are in agreement. 
This can be noticed for test cases at coordinates (1, 0.23), which represent road tracks having one curve with a maximum curvature of 0.23---an instance of a \textit{true positive} case (the corresponding cells in each map are highlighted with a solid line). Both DS\textsubscript{1} and DS\textsubscript{2} have a failure probability of 1 and, as a consequence, the \dss map also does. 
On \hdt, \msim has also a high failure probability (0.5), which confirms the high effectiveness of the \dss framework at approximating the true failure probability of \hdt.

Concerning the failure probability for \mreal---i.e., the model driving with pseudo-real driving scenes, for \davetwo and \chauffeur, \dss are better than each individual sibling in terms of correlation with \hdt. For \davetwo, DS\textsubscript{1} better predicts the failures of \hdt, while for \chauffeur, the digital siblings are better than each individual sibling. Interestingly, for \epoch, DS\textsubscript{2} better correlates with \hdt but the AUC-PRC value of \dss is the higher than the individual siblings.

\begin{tcolorbox}[colback=gray!15!white,colframe=black]
	\textbf{RQ\textsubscript{2}}: At the system-level, in four cases out of six, the failure probability of the digital siblings better correlates with the true failure probability of the digital twin w.r.t. each individual sibling. In four cases out of six, the failures obtained on the digital siblings are a better predictor of the ground truth failures experienced on the digital twin.
\end{tcolorbox}

\subsubsection{Quality of Driving (RQ\textsubscript{3})}
\autoref{table:study:rq3} shows the Pearson correlation (\textit{r}), the $p$-value, and the AUC-PRC for the comparison between DS\textsubscript{1}, DS\textsubscript{2}, \dss and \hdt, respectively. The comparison considers the correlation between the quality of driving metric experienced in DS\textsubscript{1}, DS\textsubscript{2}, \dss and the failure probability of the model on \hdt, as well as the prediction of failures from the quality of driving metric. The analysis is reported separately for both \msim (Columns 3--5) and \mreal (Columns 6--8) models. 

For \msim, the correlation between \dss and \hdt is lower than the best individual correlation for all the lane-keeping models (0.553 of \dss vs 0.621 of DS\textsubscript{1} for \davetwo, 0.792 of \dss vs 0.798 of DS\textsubscript{1} for \chauffeur, and 0.491 of \dss vs 0.511 of DS\textsubscript{1} for \epoch). For \davetwo, the \dss correlation is 22\% higher than the worst individual correlation (0.553 of \dss vs 0.429 of DS\textsubscript{2}); percentages are similar for \chauffeur and \epoch. For AUC-PRC, \dss and DS\textsubscript{1} have the same predictive power both for \davetwo and \chauffeur (i.e., respectively 0.659 and 0.940), while for \epoch the \dss prediction is slightly better than that of DS\textsubscript{1}. 
Thus, \dss mitigate the risk of relying on the testing results of a low-quality \gps (i.e., DS\textsubscript{2}). 

Concerning \mreal, we observed a similar trend, i.e., the correlation of DS\textsubscript{1} with \hdt are higher than the correlations of \dss with \hdt, although \dss always have a better correlation than the worst of the two siblings, i.e., DS\textsubscript{2}, for all lane-keeping models. The digital siblings \dss better predict the failures of \hdt for \davetwo and are equivalent to DS\textsubscript{1} for \chauffeur. For \epoch, the best predictor of the failures of \hdt is DS\textsubscript{2}, although the digital siblings are only 9\% worse.

\begin{figure*}[b]
\centering

\captionsetup{labelfont={color=myblue}}

\includegraphics[trim=1cm 0cm 46cm 0cm, clip=true, scale=0.12]
{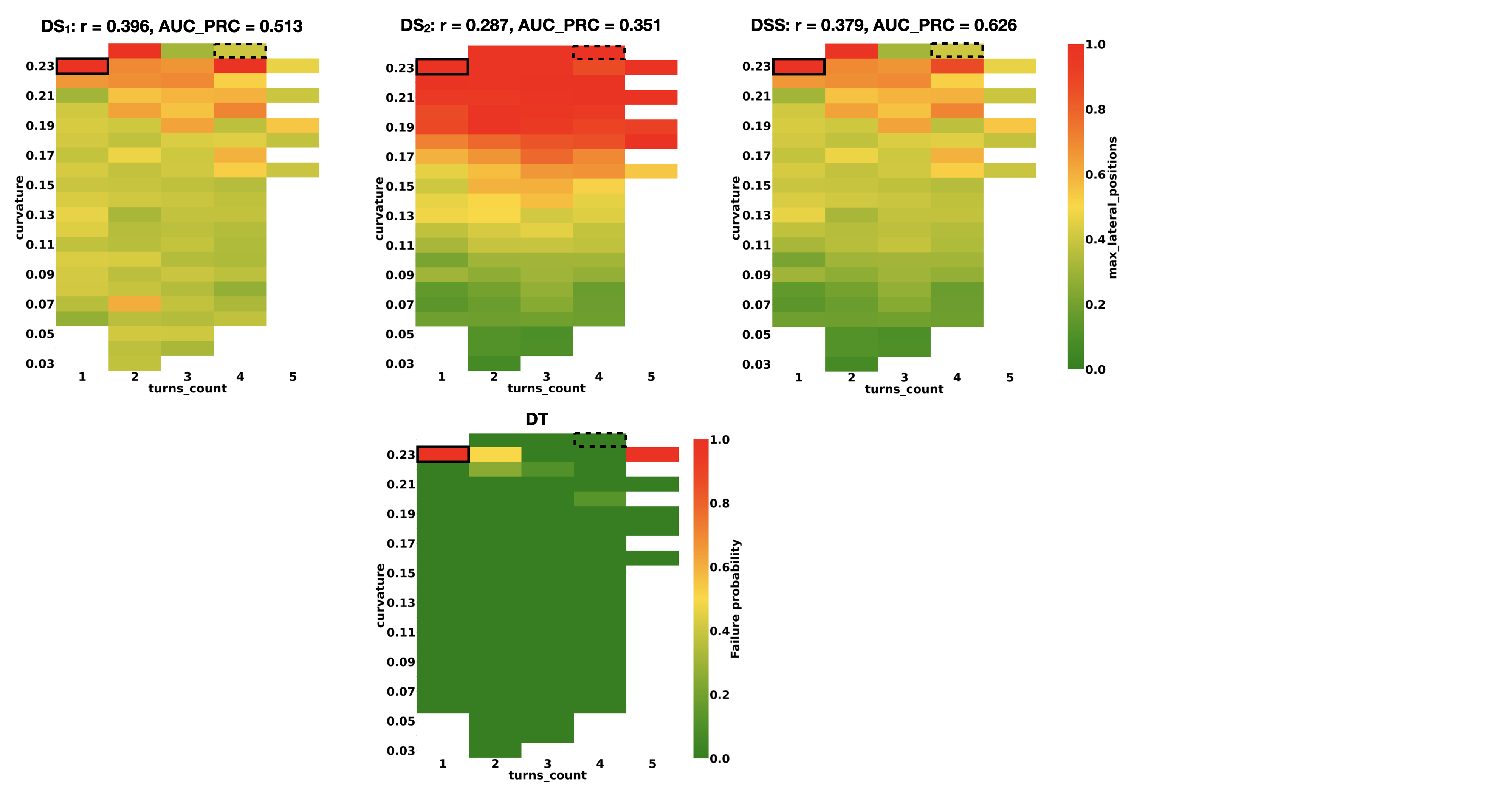}

\caption{Feature maps representing the quality of driving of \davetwo \mreal (i.e., the maximum lateral position) on the two digital siblings, DS\textsubscript{1} and DS\textsubscript{2}, their combination (\dss) and the failure probability on \hdt. Solid line cells represent a true failure predicted by \dss, while dashed line cells represent a false positive of \dstwo. Best viewed in color.} 
\label{fig:study:qm-real} 
\end{figure*}

\autoref{fig:study:qm-real} shows the four feature maps related to the quality of driving of the \mreal \davetwo model on the two digital siblings and the failure probability of \mreal on \hdt.\footnote{We report the plots for the other lane-keeping models in our replication package~\cite{tool}.} We can observe that the feature map of DS\textsubscript{1} and the feature map of \dss are similar. As a consequence, the two correlations are similar (0.396 of DS\textsubscript{1} vs 0.379 of \dss). On the other hand, the feature map of DS\textsubscript{2} is quite different from the failure probability map of \hdt, which causes the correlation to be low (0.287). 
We can observe that all siblings are able to capture the failure of the \hdt at coordinates (1, 0.23) (see the corresponding cells highlighted with a solid line). On the other hand, the test cases at coordinates (4, 0.24) triggered failures only in DS\textsubscript{2}, and \dss correctly predict that in \hdt such tests will not cause a failure.

\begin{table}[t]
	
	\captionsetup{labelfont={color=myblue}}
	\caption{\textcolor{myblue}{Results for RQ\textsubscript{3}. Bold-faced values indicate the best approach.}}
	\label{table:study:rq3}
	\setlength{\tabcolsep}{4.5pt}
	\renewcommand{\arraystretch}{1.5}
	\color{myblue}
	\centering

	\begin{tabular}{lrllllll}
		\toprule
		
		\multicolumn{1}{r}{\textbf{}} &  & \multicolumn{6}{c}{\sc Quality of Driving (RQ\textsubscript{3})} \\
		
		\cmidrule(r){3-8}
		
		\multicolumn{1}{l}{} &  
		& \multicolumn{3}{c}{\msim} 
		& \multicolumn{3}{c}{\mreal} \\
		
		\cmidrule(r){3-5} 
		\cmidrule(r){6-8} 
		
		\multicolumn{1}{l}{} &  
		& \textit{r} 
		& p-value 
		& AUC-PRC 
		
		& \textit{r} 
		& p-value 
		& AUC-PRC \\
		
		\midrule
		
		\multirow{3}{*}{\davetwo~\cite{nvidia-dave2}} & DS\textsubscript{1} vs \hdt & \textbf{0.621} & $10^{-10}$ & \textbf{0.659} & \textbf{0.396} & $10^{-4}$ & 0.513 \\
		& DS\textsubscript{2} vs \hdt & 0.429 & $10^{-5}$ & 0.496 & 0.287 & $10^{-3}$ & 0.351 \\
		& DSS vs \hdt & 0.553 & $10^{-8}$ & \textbf{0.659} & 0.379 & $10^{-4}$ & \textbf{0.626} \\

		\multicolumn{1}{l}{} &  & \multicolumn{1}{l}{} & \multicolumn{1}{l}{} & \multicolumn{1}{l}{} & \multicolumn{1}{l}{} & \multicolumn{1}{l}{} & \multicolumn{1}{l}{} \\
		
		\multirow{3}{*}{\chauffeur~\cite{chauffeur}} & DS\textsubscript{1} vs \hdt & \textbf{0.798} & $10^{-21}$ & \textbf{0.940} & \textbf{0.399} & $10^{-4}$ & \textbf{0.460} \\
		& DS\textsubscript{2} vs \hdt & 0.625 & $10^{-11}$ & 0.791 & 0.260 & 0.025 & 0.359 \\
		& DSS vs \hdt & 0.792 & $10^{-21}$ & \textbf{0.940} & 0.382 & $10^{-4}$ & \textbf{0.460} \\
		
		\multicolumn{1}{l}{} &  & \multicolumn{1}{l}{} & \multicolumn{1}{l}{} & \multicolumn{1}{l}{} & \multicolumn{1}{l}{} & \multicolumn{1}{l}{} & \multicolumn{1}{l}{} \\
		
		\multirow{3}{*}{\epoch~\cite{epoch}} & DS\textsubscript{1} vs \hdt & \textbf{0.511} & $10^{-7}$ & 0.592 & \textbf{0.554} & $10^{-8}$ & 0.608 \\
		& DS\textsubscript{2} vs \hdt & 0.355 & $10^{-4}$ & 0.541 & 0.389 & $10^{-3}$ & \textbf{0.715} \\
		& DSS vs \hdt & 0.491 & $10^{-6}$ & \textbf{0.594} & 0.529 & $10^{-7}$ & 0.651 \\

		\bottomrule
		
	\end{tabular}

\end{table}

\begin{tcolorbox}[colback=gray!15!white,colframe=black]
	\textbf{RQ\textsubscript{3}}: At the system-level, for most lane-keeping models, the quality of driving of the digital siblings correlates with the failure probability of the digital twin. This correlation is either equivalent to that of the best digital sibling or falls within the range of the two siblings. 
	In five cases out of six, the quality of driving in the digital siblings has a failure prediction capability w.r.t. the digital twin, which is equal or higher than the best individual sibling. As a result, digital siblings reduce the risk associated with relying on the least reliable simulator.
\end{tcolorbox}

\section{Discussion}\label{sec:discussion}

\head{\gpss Complementarity} When choosing candidate \gpss, our approach requires that the simulators exhibit some degree of complementarity (i.e., different physics engines), while still supporting the same encoding of test inputs. Therefore, the selected \gpss must meet the following conditions. Firstly, the simulators must be equipped with appropriate API interfaces that allow the instantiation of analogous test cases. In our context, both Udacity and BeamNG support a sequence of road points as input to instantiate the two-lane roads where the AV drives. Secondly, the simulators need to support communication with the DNN-based systems under test. In the case of a DNN-based lane-keeping AV, the simulators should be able to capture images from the vehicle's on-board camera and execute throttle steering commands to drive the vehicle. Finally, the selected  simulators should implement different physics engines. Specifically, Udacity implements soft-body dynamics, while BeamNG uses a rigid-body dynamics engine.

The worst case occurs when the two siblings disagree and the over-approximating sibling (e.g., predicting a failure) is not compensated by the under-approximating sibling (see \autoref{fig:study:fp-sim}). In most cases, we empirically observed that by predicting a failure only when there is agreement, the digital siblings are equivalent to the best of the two siblings (see RQ\textsubscript{3}). However, for the \epoch model, when considering the failure probabilities of the \mreal model, the correlation of the digital siblings is slightly worse than the worst sibling, i.e., DS\textsubscript{1} (specifically, 0.450 of \dss vs 0.469 of DS\textsubscript{2}). Despite the lowest correlation, the digital siblings have the highest capabilities of detecting the failures of \hdt.

\head{Simulated and Pseudo-real Models} We experimented with both simulated ($M_S$) and real-world models ($M_R$) as such setting is representative of the current industrial testing practices described by the NHTSA~\cite{nhtsa}. From the feature maps in \autoref{fig:study:fp-sim} and \autoref{fig:study:qm-real}, we can observe that the driving quality of $M_S$ is superior w.r.t. $M_R$ (the failure probabilities in the feature map of \hdt are higher), presumably because it is easier for a DNN to process plain artificial images from a simulator, rather than the images collected by a real-world camera during driving.

\subsection{Threats to Validity}

\subsubsection{Internal validity}
We compared all simulators under identical parameter settings.
One threat to internal validity concerns our custom implementation of DeepHyperion within the simulators. We mitigated this threat by faithfully replicating the code available in the replication package of the paper~\cite{deephyperion-tool}.
Another threat may be due to our own data collection phase and training of the lane-keeping models, which may exhibit many misbehaviors if trained inadequately. We mitigated this threat by training and fine-tuning a model which was able to drive on the majority of the training set roads consistently on all simulators.

\subsubsection{External validity}
We considered only a limited number of DNN models and simulators, which poses a threat in terms of the generalizability of our results. We tried to mitigate this threat by choosing three popular real-world DNN models, which achieved competitive scores in the Udacity challenge~\cite{udacity-challenge}. Their diversity in terms of both size and architectural structure determines different driving behaviors and increases the generalizability of our results. We considered two open-source \gpss, and we chose DonkeyCar as \hdt, as it was used as a proxy for full size self-driving cars also in previous studies~\cite{stocco-mind,2023-Stocco-EMSE,2021-01-0248,viitala,9412011}. 
Generalizability to other \gpss or \hdts would require further studies.

Our proposal focuses on testing the DNN-based lane-keeping component of an AV, by generating a large set of road scenarios. Although there are works in the literature that modify other environment objects such as weather conditions, pedestrian and other vehicles' dynamics~\cite{Abdessalem-ICSE18,samota,borg}, we chose to generate road scenarios to test the lane-keeping behavior of the DNN in isolation, avoiding the interference of other tasks, such as obstacle and pedestrian avoidance. Further studies are needed to assess the generalizability of our multi-simulator approach to driving tasks different from lane-keeping. On this regard, feature maps are a flexible tool to encode different characteristics of a test case (e.g., the intensity of the rain or the number of vehicles in the driving scenario), by adding new dimensions for each new desired feature.

\subsubsection{Construct validity}
Threats to construct validity may come from selecting  inappropriate metrics to measure the agreement of the siblings with \hdt. To address this threat we assessed such agreement from two points of view, i.e., at the model-level (RQ\textsubscript{1}), by measuring the distance between the two distributions under analysis and testing the statistical significance of the difference, and at the system-level, by measuring failure probability and quality of driving. Overall, our results show that the digital siblings are better at predicting the behavior of the lane-keeping model under test on \hdt.

\section{Related Work}\label{sec:related}

\subsection{Digital Twins for AV Testing}

Digital twins are used by researchers to reproduce real-world conditions within a simulation environment for testing purposes~\cite{10.1007/978-3-030-59155-7_39,9369807,DBLP:journals/corr/abs-2012-05841,san2021digital,9392784}. 

Yun et al.~\cite{9369807} test an object recognition system using the GTA videogame. In particular, they exploit the realism of the game engine to collect data for training an object recognition system for both collision avoidance and lane-departure prevention.
Barosan et al.~\cite{10.1007/978-3-030-59155-7_39} describe a digital twin for testing an autonomous truck. No testing was performed using the digital twin to assess the faithfulness of the simulator at reproducing real-world failures. 
Almeaibed et al.~\cite{9392784}, analyze the safety and security of digital twins and propose a general framework to address such issues during development. 
Kapteyn et al.~\cite{DBLP:journals/corr/abs-2012-05841}, propose a probabilistic graphical model to link the digital twin with its physical replica. The formal definition ensures that the calibration of the digital twin and its update with real-world data is principled and scalable. 
Similarly, San et al.~\cite{san2021digital} rely on the same mathematical tool to formalize the update of the digital twin with the goal of using it throughout the whole lifecycle of its physical replica, i.e., from the design to the operation phase.
Veledar et al.~\cite{veledar2019digital} propose a multi-metrics approach for security and safety validation for the design of a digital twin for autonomous driving.

Such works mostly focus on the design of the digital twin and its update during the development of the physical replica.
Differently, in our paper we investigate testing transferability between digital siblings, i.e., a multi-simulator approach considering both simulated and pseudo-real images as input to the DNN.
 
\subsection{Empirical Studies}

Simulation platforms are often decoupled from the real world complexities~\cite{icst-survey-robotics}, which confirmed the need for real-world testing of cyber-physical systems. 
Our work is the first to propose the usage of a multi-simulator approach, called digital siblings, to mitigate the fidelity gap in the field of autonomous driving testing. 

Concerning comparative studies across simulators, to the best of our knowledge, the only study that empirically compares the same AV on different simulation platforms is by Borg et al.~\cite{borg}. The authors investigate the use of multiple \gps for testing a pedestrian vision detection system. The study compares a large set of test scenarios on both PreScan~\cite{prescan} and Pro-SiVIC~\cite{pro-sivic} and reports  low agreement between testing results across the two simulation platforms. No assessment is performed of their correlation with a digital twin or a physical vehicle. In our paper, we take a step ahead, and we show how the (dis)agreements can be leveraged to mitigate the fidelity gap: by combining the predictions of two general-purpose simulators we successfully covered the gap with a DT for a scaled physical vehicle. In another work, Amini et al.~\cite{amini2023evaluating} evaluates the degree of flakiness affecting two widely-used open-source AV simulators and five diverse test setups, showing that test flakiness in AV is a common issue and can significantly impact the test results obtained by randomized algorithms.

Other studies compare model-level vs system-level testing metrics within a simulation environment~\cite{briand-offline-emse}.
In our empirical work, we focused on the difference between general-purpose and digital twin driving simulators. We use offline and online testing to measure the gap between single- and multi-simulator approaches at approximating a digital twin, a previously unexplored topic. Our proposition is also meant to prevent the flakiness occurring within a single simulation platform, by relying on an ensemble of simulators.

\subsection{AV Testing Approaches}

Most approaches use \textit{model-level testing} (i.e., offline testing of single image predictions) to test DNN autopilots under corrupted images~\cite{deeptest,physigan} or GAN-generated driving scenarios~\cite{deeproad}, without however testing the self-driving software in its operational domain.
In our work, we assess the effectiveness of our digital siblings with model-level testing in terms of prediction error distributions, but we also consider online testing at the system-level.  

Another model-level testing approach is by Talwar et al.~\cite{sim-to-real-lgsvl}. Their focus is to test the generalizability on real-world data of multiple object detection models trained on simulated images. On the other hand, we use an Image-to-Image translation architecture~\cite{cyclegan} to translate simulated images into real-world images both to evaluate the lane-keeping model offline and to test it online at the system-level.

Concerning \textit{system-level testing} for AVs, researchers proposed techniques to generate scenarios that cause AVs to misbehave~\cite{2020-Stocco-ICSE,asfault,2021-Stocco-JSEP,2022-Stocco-ASE,arxiv.2203.12026,deeproad,drivefuzz,zhongETAL2021,liETAL2020,Jha2019MLBasedFI,10.1145/3597926.3598072,2024-Grewal-ICST}. Among the existing test generators, in this work we adopted DeepHyperion by Zohdinasab et al.~\cite{deephyperion}, a tool that uses illumination search to extensively cover a map of structural input features, which allowed us to easily group identical or equivalent failure conditions occurring in the same feature map cell. 
Ul Haq et al.~\cite{samota} use ML regressors as surrogate models to mimic the simulator's outcome. 

These works only consider single-simulator approaches to testing. Their generalizability to a multi-simulator approach, such as the digital siblings proposed in this paper, or to cross-simulator testing, is overlooked in the existing literature. 

\section{Conclusions and Future Work}\label{sec:conclusions}

In this paper, we propose a multi-simulator approach named digital siblings, to improve simulation-based testing of the lane-keeping component of an autonomous vehicle.
In our approach, we test the autonomous driving software by generating road scenarios in two general-purpose simulators, to better approximate the behavior of the lane-keeping model on a digital twin. We combine the testing outputs of the model on the two simulators in a conservative way, giving priority to the agreements on possible failures, where it is more likely to observe the same failing behavior on the digital twin.

At the model level, our results show that the digital siblings approximate the model predictions on the digital twin better than each individual simulator. At the system-level, the digital siblings are able to predict the failures of the model on the digital twin better than each single simulator.

In our future work we plan to extend our case study to more than two general-purpose simulators, and to study different ways to combine them based on the characteristics of each simulator and those of the digital twin.

\section{Acknowledgments}

We thank BeamNG GmbH for providing us the license for the driving simulator.

\section{Declarations}

\subsection{Funding and/or Conflicts of interests/Competing interests}
This work was partially supported by the H2020 project PRECRIME, funded under the ERC Advanced Grant 2017 Program (ERC Grant Agreement n. 787703).
The authors declared that they have no conflict of interest.

\subsection{Data Availability}
The software artifacts and our results are publicly available~\cite{tool}.

\subsection{Version of Record}
This version of the contribution has been accepted for publication, after peer review (when applicable) but is not the Version of Record and does not reflect post-acceptance improvements, or any corrections. The Version of Record is available online at: \href{https://doi.org/10.1007/s10664-024-10458-4}{https://doi.org/10.1007/s10664-024-10458-4}. Use of this Accepted Version is subject to the publisher’s Accepted Manuscript terms of use \href{https://www.springernature.com/gp/open-research/policies/accepted-manuscript-terms}{https://www.springernature.com/gp/open-research/policies/accepted-manuscript-terms}.

\balance
\bibliographystyle{spmpsci}
\bibliography{paper}

\begin{thebibliography}{10}
\providecommand{\url}[1]{{#1}}
\providecommand{\urlprefix}{URL }
\expandafter\ifx\csname urlstyle\endcsname\relax
  \providecommand{\doi}[1]{DOI~\discretionary{}{}{}#1}\else
  \providecommand{\doi}{DOI~\discretionary{}{}{}\begingroup
  \urlstyle{rm}\Url}\fi

\bibitem{icst-survey-robotics}
Afzal, A., Katz, D.S., Le~Goues, C., Timperley, C.S.: Simulation for robotics
  test automation: Developer perspectives.
\newblock In: 2021 14th IEEE Conference on Software Testing, Verification and
  Validation (ICST), pp. 263--274. IEEE (2021)

\bibitem{9392784}
Almeaibed, S., Al-Rubaye, S., Tsourdos, A., Avdelidis, N.P.: Digital twin
  analysis to promote safety and security in autonomous vehicles.
\newblock IEEE Communications Standards Magazine \textbf{5}(1), 40--46 (2021).
\newblock \doi{10.1109/MCOMSTD.011.2100004}

\bibitem{amini2023evaluating}
Amini, M.H., Naseri, S., Nejati, S.: Evaluating the impact of flaky simulators
  on testing autonomous driving systems (2023)

\bibitem{wgan}
Arjovsky, M., Chintala, S., Bottou, L.: Wasserstein generative adversarial
  networks.
\newblock In: International conference on machine learning, pp. 214--223. PMLR
  (2017)

\bibitem{10.1007/978-3-030-59155-7_39}
Barosan, I., Basmenj, A.A., Chouhan, S.G.R., Manrique, D.: Development of a
  virtual simulation environment and a digital twin of an autonomous driving
  truck for a distribution center.
\newblock In: Software Architecture, pp. 542--557. Springer, Cham (2020)

\bibitem{10.1145/378456.378511}
Barry, P.J., Goldman, R.N.: A recursive evaluation algorithm for a class of
  catmull-rom splines.
\newblock SIGGRAPH Comput. Graph.  (1988)

\bibitem{beamng}
BeamNG.research: {BeamNG GmbH}.
\newblock \url{ https://www.beamng.gmbh/research} (2022)

\bibitem{Abdessalem-ICSE18}
{Ben Abdessalem}, R., {Nejati}, S., {C. Briand}, L., {Stifter}, T.: Testing
  vision-based control systems using learnable evolutionary algorithms.
\newblock In: 2018 IEEE/ACM 40th International Conference on Software
  Engineering (ICSE) (2018)

\bibitem{wayve-sim2real}
Bewley, A., Rigley, J., Liu, Y., Hawke, J., Shen, R., Lam, V.D., Kendall, A.:
  Learning to drive from simulation without real world labels.
\newblock In: 2019 International conference on robotics and automation (ICRA),
  pp. 4818--4824. IEEE (2019)

\bibitem{waymo}
BGR~Media, L.: Waymo’s self-driving cars hit 10 million miles.
\newblock
  \url{https://techcrunch.com/2018/10/10/waymos-self-driving-cars-hit-10-million-miles}
  (2018)

\bibitem{sbst2023}
Biagiola, M., Klikovits, S., Peltomaki, J., Riccio, V.: Sbft tool competition
  2023-cyber-physical systems track.
\newblock In: 16th IEEE/ACM International Workshop on Search-Based And Fuzz
  Testing, SBFT (2023)

\bibitem{nvidia-dave2}
Bojarski, M., Del~Testa, D., Dworakowski, D., Firner, B., Flepp, B., Goyal, P.,
  Jackel, L.D., Monfort, M., Muller, U., Zhang, J., Zhang, X., Zhao, J., Zieba,
  K.: End to end learning for self-driving cars.
\newblock CoRR \textbf{abs/1604.07316} (2016)

\bibitem{borg}
Borg, M., Abdessalem, R.B., Nejati, S., Jegeden, F.X., Shin, D.: Digital twins
  are not monozygotic--cross-replicating adas testing in two industry-grade
  automotive simulators.
\newblock In: ICST '21. IEEE (2021)

\bibitem{pros-and-cons-gans}
Borji, A.: Pros and cons of gan evaluation measures.
\newblock Computer Vision and Image Understanding \textbf{179}, 41--65 (2019)

\bibitem{10.5555/2981562.2981583}
Bottou, L., Bousquet, O.: The tradeoffs of large scale learning.
\newblock In: Proceedings of NIPS '07 (2007)

\bibitem{ad-market}
Boutan, E.: Autonomous driving market overview.
\newblock
  \url{https://medium.com/swlh/autonomous-driving-market-overview-b8c71d81c072}
  (2020)

\bibitem{comprehensive-sfc-test}
Cerf, V.G.: A comprehensive self-driving car test.
\newblock Communications of the ACM \textbf{61}(2) (2018)

\bibitem{chauffeur}
{Team Chauffeur, ``Steering angle model: Chauffeur.''}.
\newblock
  \url{https://github.com/udacity/self-driving-car/tree/master/steering-models/community-models/chauffeur}
  (2016)

\bibitem{epoch}
{Team Epoch, ``Steering angle model: Epoch.''}.
\newblock
  \url{https://github.com/udacity/self-driving-car/tree/master/steering-models/community-models/cg23}
  (2016)

\bibitem{10.1145/3597926.3598072}
Cheng, M., Zhou, Y., Xie, X.: Behavexplor: Behavior diversity guided testing
  for autonomous driving systems.
\newblock In: Proceedings of the 32nd ACM SIGSOFT International Symposium on
  Software Testing and Analysis, ISSTA 2023, p. 488–500. Association for
  Computing Machinery, New York, NY, USA (2023).
\newblock \doi{10.1145/3597926.3598072}.
\newblock \urlprefix\url{https://doi.org/10.1145/3597926.3598072}

\bibitem{non-parametric-book}
Conover, W.J.: Practical nonparametric statistics, vol. 350.
\newblock john wiley \& sons (1999)

\bibitem{deephyperion-tool}
{DeepHyperion Replication package}.
\newblock \url{https://github.com/testingautomated-usi/DeepHyperion} (2022)

\bibitem{donkeycar}
{Donkey Car}.
\newblock \url{https://www.donkeycar.com/} (2021)

\bibitem{farag2020complex}
Farag, W.: Complex trajectory tracking using pid control for autonomous
  driving.
\newblock International Journal of Intelligent Transportation Systems Research
  \textbf{18}(2), 356--366 (2020)

\bibitem{evosuite}
Fraser, G., Arcuri, A.: Whole test suite generation.
\newblock IEEE Transactions on Software Engineering \textbf{39}(2), 276--291
  (2012)

\bibitem{sbst2022}
Gambi, A., Jahangirova, G., Riccio, V., Zampetti, F.: {SBST} tool competition
  2022.
\newblock In: 2022 IEEE/ACM 15th International Workshop on Search-Based
  Software Testing (SBST), pp. 25--32. IEEE (2022)

\bibitem{gambi-beamng}
Gambi, A., Maul, P., Mueller, M., Stamatogiannakis, L., Fischer, T.,
  Panichella, S.: Soft-body simulation and procedural generation for the
  development and testing of cyber-physical systems.
\newblock Tech. rep., BeamNG (2019)

\bibitem{asfault}
Gambi, A., Mueller, M., Fraser, G.: Automatically testing self-driving cars
  with search-based procedural content generation.
\newblock In: Proceedings of ISSTA '19 (2019)

\bibitem{fse-survey-robotics}
Garc{\'\i}a, S., Str{\"u}ber, D., Brugali, D., Berger, T., Pelliccione, P.:
  Robotics software engineering: A perspective from the service robotics
  domain.
\newblock In: Proceedings of ESEC/FSE '20, pp. 593--604 (2020)

\bibitem{2024-Grewal-ICST}
Grewal, R., Tonella, P., Stocco, A.: {Predicting Safety Misbehaviours in
  Autonomous Driving Systems using Uncertainty Quantification} p. 12 pages
  (2024)

\bibitem{grigorescu2020survey}
Grigorescu, S., Trasnea, B., Cocias, T., Macesanu, G.: A survey of deep
  learning techniques for autonomous driving.
\newblock Journal of Field Robotics \textbf{37}(3), 362--386 (2020)

\bibitem{pro-sivic}
Group, E.: Esi prosivic.
\newblock
  \url{https://myesi.esi-group.com/downloads/software-downloads/pro-sivic-2021.0}
  (2021)

\bibitem{samota}
Haq, F.U., Shin, D., Briand, L.C.: Efficient online testing for dnn-enabled
  systems using surrogate-assisted and many-objective optimization.
\newblock In: 44th {IEEE/ACM} 44th International Conference on Software
  Engineering, {ICSE} 2022, Pittsburgh, PA, USA, May 25-27, 2022, pp. 811--822.
  {ACM} (2022).
\newblock \doi{10.1145/3510003.3510188}.
\newblock \urlprefix\url{https://doi.org/10.1145/3510003.3510188}

\bibitem{briand-offline-emse}
Haq, F.U., Shin, D., Nejati, S., Briand, L.: Can offline testing of deep neural
  networks replace their online testing?
\newblock Empirical Software Engineering  (2021)

\bibitem{hu2023sim2real}
Hu, X., Li, S., Huang, T., Tang, B., Chen, L.: Sim2real and digital twins in
  autonomous driving: A survey (2023)

\bibitem{2021-Jahangirova-ICST}
Jahangirova, G., Stocco, A., Tonella, P.: Quality metrics and oracles for
  autonomous vehicles testing.
\newblock In: Proceedings of 14th IEEE International Conference on Software
  Testing, Verification and Validation, ICST '21. IEEE (2021)

\bibitem{Jha2019MLBasedFI}
Jha, S., Banerjee, S.S., Tsai, T., Hari, S.K.S., Sullivan, M.B., Kalbarczyk,
  Z.T., Keckler, S.W., Iyer, R.K.: Ml-based fault injection for autonomous
  vehicles: A case for bayesian fault injection.
\newblock 2019 49th Annual IEEE/IFIP International Conference on Dependable
  Systems and Networks (DSN) pp. 112--124 (2019).
\newblock \urlprefix\url{https://api.semanticscholar.org/CorpusID:195776612}

\bibitem{DBLP:journals/corr/abs-2012-05841}
Kapteyn, M.G., Pretorius, J.V.R., Willcox, K.E.: A probabilistic graphical
  model foundation for enabling predictive digital twins at scale.
\newblock CoRR \textbf{abs/2012.05841} (2020)

\bibitem{DBLP:journals/corr/abs-2101-05337}
Kaur, P., Taghavi, S., Tian, Z., Shi, W.: A survey on simulators for testing
  self-driving cars.
\newblock CoRR \textbf{abs/2101.05337} (2021).
\newblock \urlprefix\url{https://arxiv.org/abs/2101.05337}

\bibitem{drivefuzz}
Kim, S., Liu, M., Rhee, J.J., Jeon, Y., Kwon, Y., Kim, C.H.: {DriveFuzz}.
\newblock In: Proceedings of the 2022 {ACM} {SIGSAC} Conference on Computer and
  Communications Security. {ACM} (2022).
\newblock \doi{10.1145/3548606.3560558}.
\newblock \urlprefix\url{https://doi.org/10.1145%2F3548606.3560558}

\bibitem{kingma2014adam}
Kingma, D.P., Ba, J.: Adam: A method for stochastic optimization.
\newblock arXiv preprint arXiv:1412.6980  (2014)

\bibitem{physigan}
Kong, Z., Guo, J., Li, A., Liu, C.: Physgan: Generating
  physical-world-resilient adversarial examples for autonomous driving.
\newblock In: Proceedings of the IEEE/CVF Conference on Computer Vision and
  Pattern Recognition, pp. 14,254--14,263 (2020)

\bibitem{challenges-av-testing}
Koopman, P., Wagner, M.: Challenges in autonomous vehicle testing and
  validation.
\newblock SAE International Journal of Transportation Safety  (2016)

\bibitem{siemens}
Kothlow, C.: The power of a multi-purpose digital twin.
\newblock
  \url{https://blogs.sw.siemens.com/simcenter/the-power-of-a-multi-purpose-digital-twin/}
  (2021)

\bibitem{2024-Lambertenghi-ICST}
Lambertenghi, S.C., Stocco, A.: {Assessing Quality Metrics for Neural Reality
  Gap Input Mitigation in Autonomous Driving Testing} p. 12 pages (2024)

\bibitem{liETAL2020}
Li, G., Li, Y., Jha, S., Tsai, T., Sullivan, M., Hari, S.K.S., Kalbarczyk, Z.,
  Iyer, R.: Av-fuzzer: Finding safety violations in autonomous driving systems.
\newblock In: 2020 IEEE 31st International Symposium on Software Reliability
  Engineering (ISSRE), pp. 25--36 (2020).
\newblock \doi{10.1109/ISSRE5003.2020.00012}

\bibitem{pynguin}
Lukasczyk, S., Kroi{\ss}, F., Fraser, G.: Automated unit test generation for
  python.
\newblock In: International Symposium on Search Based Software Engineering, pp.
  9--24. Springer (2020)

\bibitem{essentials}
Luke, S.: Essentials of metaheuristics, vol.~2.
\newblock Lulu Raleigh (2013)

\bibitem{outsourcing-av-development}
May, C.: Why automotive companies outsource software development services.
\newblock
  \url{https://medium.datadriveninvestor.com/why-automotive-companies-outsource-software-development-services-54a806458b4?gi=9d9b4f45e9ba}
  (2019)

\bibitem{arxiv.2203.12026}
Moghadam, M.H., Borg, M., Saadatmand, M., Mousavirad, S.J., Bohlin, M., Lisper,
  B.: Machine learning testing in an adas case study using
  simulation-integrated bio-inspired search-based testing (2022)

\bibitem{mapelites}
Mouret, J.B., Clune, J.: Illuminating search spaces by mapping elites.
\newblock arXiv preprint arXiv:1504.04909  (2015)

\bibitem{dynamosa}
Panichella, A., Kifetew, F.M., Tonella, P.: Automated test case generation as a
  many-objective optimisation problem with dynamic selection of the targets.
\newblock IEEE Transactions on Software Engineering \textbf{44}(2), 122--158
  (2017)

\bibitem{sbst2021}
Panichella, S., Gambi, A., Zampetti, F., Riccio, V.: {SBST} tool competition
  2021.
\newblock In: 2021 IEEE/ACM 14th International Workshop on Search-Based
  Software Testing (SBST), pp. 20--27. IEEE (2021)

\bibitem{PhysX}
{Nvidia PhysX}.
\newblock \url{https://developer.nvidia.com/physx-sdk} (2022)

\bibitem{wasserstein}
Ramdas, A., Garc{\'\i}a~Trillos, N., Cuturi, M.: On wasserstein two-sample
  testing and related families of nonparametric tests.
\newblock Entropy \textbf{19}(2), 47 (2017)

\bibitem{tool}
{Replication package}.
\newblock \url{https://github.com/testingautomated-usi/maxitwo} (2023)

\bibitem{deepjanus}
Riccio, V., Tonella, P.: Model-based exploration of the frontier of behaviours
  for deep learning system testing.
\newblock In: Proceedings of ESEC/FSE (2020)

\bibitem{s19030648}
Rosique, F., Navarro, P.J., Fernández, C., Padilla, A.: A systematic review of
  perception system and simulators for autonomous vehicles research.
\newblock Sensors \textbf{19}(3) (2019).
\newblock \doi{10.3390/s19030648}

\bibitem{saadonline}
Saad, D.: Online algorithms and stochastic approximations.
\newblock Online Learning  (1998)

\bibitem{prc-justification}
Saito, T., Rehmsmeier, M.: The precision-recall plot is more informative than
  the roc plot when evaluating binary classifiers on imbalanced datasets.
\newblock PloS one \textbf{10}(3), e0118,432 (2015)

\bibitem{san2021digital}
San, O.: The digital twin revolution.
\newblock Nature Computational Science \textbf{1}(5), 307--308 (2021)

\bibitem{prescan}
Software, S.D.I.: Simcenter prescan.
\newblock
  \url{https://www.plm.automation.siemens.com/global/en/products/simcenter/prescan.html}
  (2022)

\bibitem{2022-Stocco-ASE}
Stocco, A., Nunes, P.J., d'Amorim, M., Tonella, P.: Thirdeye: Attention maps
  for safe autonomous driving systems.
\newblock In: Proceedings of 37th IEEE/ACM International Conference on
  Automated Software Engineering, ASE '22. IEEE/ACM (2022)

\bibitem{stocco-mind}
Stocco, A., Pulfer, B., Tonella, P.: {Mind the Gap! A Study on the
  Transferability of Virtual vs Physical-world Testing of Autonomous Driving
  Systems}.
\newblock IEEE Transactions on Software Engineering  (2022).
\newblock \urlprefix\url{https://ieeexplore.ieee.org/document/9869302}

\bibitem{2023-Stocco-EMSE}
Stocco, A., Pulfer, B., Tonella, P.: {Model vs System Level Testing of
  Autonomous Driving Systems: A Replication and Extension Study}.
\newblock Empirical Software Engineering  (2023)

\bibitem{2020-Stocco-GAUSS}
Stocco, A., Tonella, P.: Towards anomaly detectors that learn continuously.
\newblock In: Proceedings of 31st International Symposium on Software
  Reliability Engineering Workshops, ISSREW 2020. IEEE (2020)

\bibitem{2021-Stocco-JSEP}
Stocco, A., Tonella, P.: Confidence-driven weighted retraining for predicting
  safety-critical failures in autonomous driving systems.
\newblock Journal of Software: Evolution and Process  (2021).
\newblock \doi{10.1002/smr.2386}

\bibitem{2020-Stocco-ICSE}
Stocco, A., Weiss, M., Calzana, M., Tonella, P.: Misbehaviour prediction for
  autonomous driving systems.
\newblock In: Proceedings of 42nd International Conference on Software
  Engineering, ICSE '20. ACM (2020)

\bibitem{sim-to-real-lgsvl}
Talwar, D., Guruswamy, S., Ravipati, N., Eirinaki, M.: Evaluating validity of
  synthetic data in perception tasks for autonomous vehicles.
\newblock In: 2020 IEEE International Conference On Artificial Intelligence
  Testing (AITest), pp. 73--80. IEEE (2020)

\bibitem{survey-lei-ma}
Tang, S., Zhang, Z., Zhang, Y., Zhou, J., Guo, Y., Liu, S., Guo, S., Li, Y.,
  Ma, L., Xue, Y., Liu, Y.: A survey on automated driving system testing:
  Landscapes and trends.
\newblock CoRR \textbf{abs/2206.05961} (2022).
\newblock \doi{10.48550/arXiv.2206.05961}.
\newblock \urlprefix\url{https://doi.org/10.48550/arXiv.2206.05961}

\bibitem{donkey}
Tawn~Kramer, M.E., contributors: Donkeycar.
\newblock \url{https://www.donkeycar.com/} (2022)

\bibitem{udacity-sim}
Team, U.: Udacity's self-driving car simulator.
\newblock \url{https://github.com/tsigalko18/self-driving-car-sim} (2019)

\bibitem{udacity-challenge}
Team, U.: Udacity self-driving car challenge.
\newblock \url{https://github.com/udacity/self-driving-car/} (2020)

\bibitem{deeptest}
Tian, Y., Pei, K., Jana, S., Ray, B.: Deeptest: Automated testing of
  deep-neural-network-driven autonomous cars.
\newblock In: Proceedings of ICSE '18. ACM (2018)

\bibitem{precrashreport}
U.S. Department~of Transportation, N.H.T.S.A.: Pre-crash scenario typology for
  crash avoidance research (2007)

\bibitem{nhtsa}
of~Transportation, U.D.: A framework for automated driving system testable
  cases and scenarios.
\newblock \url{https://rosap.ntl.bts.gov/view/dot/38824/dot_38824_DS1.pdf}
  (2018)

\bibitem{udacity-simulator}
{Udacity}: {A self-driving car simulator built with Unity}.
\newblock \url{https://github.com/udacity/self-driving-car-sim} (2017).
\newblock Online; accessed 18 August 2019

\bibitem{unity}
Unity3d.
\newblock \url{https://unity.com} (2021)

\bibitem{cagatay}
{van Dinter}, R., Tekinerdogan, B., Catal, C.: Predictive maintenance using
  digital twins: A systematic literature review.
\newblock Information and Software Technology  (2022)

\bibitem{veledar2019digital}
Veledar, O., Damjanovic-Behrendt, V., Macher, G.: Digital twins for
  dependability improvement of autonomous driving.
\newblock In: Systems, Software and Services Process Improvement: 26th European
  Conference, EuroSPI 2019, Edinburgh, UK, September 18--20, 2019, Proceedings
  26, pp. 415--426. Springer (2019)

\bibitem{2021-01-0248}
Verma, A., Bagkar, S., Allam, N.V.S., Raman, A., Schmid, M., Krovi, V.N.:
  {Implementation and Validation of Behavior Cloning Using Scaled Vehicles}.
\newblock In: SAE WCX Digital Summit. SAE International (2021).
\newblock \doi{https://doi.org/10.4271/2021-01-0248}

\bibitem{viitala}
Viitala, A., Boney, R., Kannala, J.: {Learning to Drive Small Scale Cars from
  Scratch}.
\newblock CoRR \textbf{abs/2008.00715} (2020).
\newblock \urlprefix\url{https://arxiv.org/abs/2008.00715}

\bibitem{waabi-world}
Waabi: Waabi world.
\newblock \url{https://waabi.ai/waabi-world/} (2022)

\bibitem{simulation-city}
Waymo: Simulation city.
\newblock \url{https://waymo.com/blog/2021/06/SimulationCity.html} (2021)

\bibitem{wayve-infinity}
Wayve: Introducing wayve infinity simulator.
\newblock \url{https://wayve.ai/blog/introducing-wayve-infinity-simulator/}
  (2022)

\bibitem{9369807}
Yun, H., Park, D.: Simulation of self-driving system by implementing digital
  twin with gta5.
\newblock In: 2021 International Conference on Electronics, Information, and
  Communication (ICEIC), pp. 1--2 (2021).
\newblock \doi{10.1109/ICEIC51217.2021.9369807}

\bibitem{yurtsever2020survey}
Yurtsever, E., Lambert, J., Carballo, A., Takeda, K.: A survey of autonomous
  driving: Common practices and emerging technologies.
\newblock IEEE access \textbf{8}, 58,443--58,469 (2020)

\bibitem{deeproad}
Zhang, M., Zhang, Y., Zhang, L., Liu, C., Khurshid, S.: Deeproad: Gan-based
  metamorphic testing and input validation framework for autonomous driving
  systems.
\newblock In: Proceedings of ASE '18 (2018)

\bibitem{zhongETAL2021}
Zhong, Z., Kaiser, G., Ray, B.: Neural network guided evolutionary fuzzing for
  finding traffic violations of autonomous vehicles (2021)

\bibitem{9412011}
Zhou, H., Chen, X., Zhang, G., Zhou, W.: {Deep Reinforcement Learning for
  Autonomous Driving by Transferring Visual Features}.
\newblock In: 2020 25th International Conference on Pattern Recognition (ICPR)
  (2021).
\newblock \doi{10.1109/ICPR48806.2021.9412011}

\bibitem{cyclegan}
Zhu, J.Y., Park, T., Isola, P., Efros, A.A.: Unpaired image-to-image
  translation using cycle-consistent adversarial networks.
\newblock In: Computer Vision (ICCV), 2017 IEEE International Conference on
  (2017)

\bibitem{zohdinasabefficient}
Zohdinasab, T., Riccio, V., Gambi, A., Tonella, P.: Efficient and effective
  feature space exploration for testing deep learning systems.
\newblock ACM Transactions on Software Engineering and Methodology

\bibitem{deephyperion}
Zohdinasab, T., Riccio, V., Gambi, A., Tonella, P.: Deephyperion: exploring the
  feature space of deep learning-based systems through illumination search.
\newblock In: Proceedings of ISSTA '21 (2021)

\end{thebibliography}

\end{document}